%% file: main.tex
\documentclass[11pt]{article}
\usepackage{enumitem}

\usepackage[letterpaper,margin=1in]{geometry}
\usepackage[square,comma]{natbib}
\usepackage{authblk}
\usepackage{etoolbox}
\makeatletter\patchcmd{\AB@maketitle}{\rlap}{\mbox}{}{}\makeatother 
\usepackage{multirow}
\usepackage{makecell}

\usepackage[utf8]{inputenc}
\usepackage[T1]{fontenc}
\usepackage[tt=false]{libertine}
\usepackage{hyperref}
\usepackage{url}
\usepackage{booktabs}
\usepackage{amsfonts}
\usepackage{nicefrac}
\usepackage{microtype}
\usepackage{xcolor}
\usepackage{booktabs}
\usepackage[table, dvipsnames]{xcolor}
\usepackage{xargs}
\usepackage{framed}
\usepackage{optidef}
\usepackage{amsthm}
\usepackage{amsmath}
\usepackage{cleveref}
\usepackage{comment}
\usepackage[colorinlistoftodos,prependcaption,textsize=small]{todonotes}

\usepackage{booktabs}
\usepackage{siunitx}

\def\tmp#1#2#3{%
  \definecolor{Hy#1color}{#2}{#3}%
  \hypersetup{#1color=Hy#1color}}
\tmp{link}{HTML}{800006}
\tmp{cite}{HTML}{2E7E2A}
\tmp{file}{HTML}{131877}
\tmp{url} {HTML}{8A0087}
\tmp{menu}{HTML}{727500}
\tmp{run} {HTML}{137776}
\def\tmp#1#2{%
  \colorlet{Hy#1bordercolor}{Hy#1color#2}%
  \hypersetup{colorlinks, #1bordercolor=Hy#1bordercolor}}
\tmp{link}{!60!white}
\tmp{cite}{!60!white}
\tmp{file}{!60!white}
\tmp{url} {!60!white}
\tmp{menu}{!60!white}
\tmp{run} {!60!white}
\usepackage{thm-restate}

\newcommand\numberthis{\addtocounter{equation}{1}\tag{\theequation}}

\newtheorem{claim}{Claim}

\makeatletter
\def\munderbar#1{\underline{\sbox\tw@{$#1$}\dp\tw@\z@\box\tw@}}
\makeatother

\newtheorem{proposition}{Proposition}
\newtheorem{theorem}{Theorem}
\newtheorem{lemma}{Lemma}

\newtheorem{definition}{Definition}

\newtheorem{corollary}{Corollary}

\newtheorem{assumption}{Assumption}

\usepackage{ltablex}

\newcommand{\cor}{Q}

\newcommand{\PBE}{\mathrm{PE}}
\newcommand{\EFBCE}{\mathrm{EFCE}}
\newcommand{\EFBCCE}{\mathrm{EFCCE}}

\newcommand{\RO}{\emph{Riley outcome}}

\newcommand{\SPBE}{\mathcal{S}^{\PBE}}
\newcommand{\SEFBCE}{\mathcal{S}^{\EFBCE}}
\newcommand{\SEFBCCE}{\mathcal{S}^{\EFBCCE}}

\usetikzlibrary{calc}

\DeclareMathOperator{\supp}{supp}

\title{On the Convergence of First-Order Methods in Signaling Games\thanks{Accepted at the 40th Conference on Neural Information Processing Systems (NeurIPS 2026).}}

\author{Rilind Sahitaj\thanks{Equal contribution.}}
\author{Artem Tsikiridis\protect\footnotemark[2]}
\author{Martin Bichler}
\affil{Department of Computer Science, Technical University of Munich\\
\texttt{\{rilind.sahitaj, artem.tsikiridis, m.bichler\}@tum.de}}
\date{}

\begin{document}

\maketitle

\begin{abstract}
We study first-order learning dynamics in Bayesian signaling games, a minimal extensive-form model of strategic interaction under asymmetric information. In these games, a Sender observes a private type and chooses a signal, after which a Receiver forms a posterior belief and selects an action. This Bayesian belief update makes the induced learning field fundamentally different from the affine game-gradient fields of normal-form games. Empirically, we find a sharp dichotomy: standard online learning algorithms reliably converge to strict Perfect Bayesian Equilibria (PE), but not to non-strict ones. We show that this behavior cannot be explained by the global geometric conditions commonly used to prove last-iterate convergence, such as monotonicity or the Minty variational inequality. Indeed, these conditions can fail even in binary signaling games. Instead, we identify the relevant local geometry. Our main result characterizes variational stability in finite signaling games: a PE is variationally stable if and only if it is strict. This yields local last-iterate convergence guarantees for online mirror descent and regularized dual averaging near every strict PE. We then study global average-iterate convergence and identify a class of signaling games in which coarse-trigger regret-minimizing dynamics converge to the unique PE. Finally, we complement the theory with experiments comparing projected gradient ascent, mirror-based methods, optimistic variants, and PPO across canonical signaling-game families. The results position signaling games as a compact testbed for understanding how Bayesian belief formation reshapes the convergence geometry of multi-agent learning.
\end{abstract}

\section{Introduction}

A central objective in learning in games is to understand when online learning
algorithms converge to equilibrium. In normal-form games, many of the strongest
convergence guarantees are obtained by imposing geometric conditions on the
game-gradient field, such as monotonicity \citep{facchinei2003finite}, the Minty
variational inequality \citep{mertikopoulos2018cycles}, and variational stability
\citep{mertikopoulos2019learning}. These conditions connect local payoff
improvements to global or local convergence guarantees for first-order learning
dynamics, including online mirror descent \citep{zhou2017mirror}, regularized
dual averaging \citep{mertikopoulos2019learning}, extragradient methods
\citep{mokhtari2020unified}, and optimistic variants
\citep{daskalakis2018training}. A key reason why this theory is tractable in
finite normal-form games is that expected payoffs are multilinear in mixed
strategies, so the induced game-gradient field is affine on the mixed extension
of the game. Much less is known about whether this geometric perspective extends
to extensive-form games with asymmetric information, where beliefs and
strategies interact through Bayesian updating.

We study this question in finite Bayesian signaling games, the canonical
extensive-form model of strategic interaction under asymmetric information
\citep{spence1978job,crawford1982strategic,sobel2020signaling}. A Sender observes
a private type and chooses a signal; a Receiver observes the signal, forms a
posterior belief over types, and then chooses an action. Signaling games are a
minimal setting in which first-order learning must interact with endogenous
belief formation. The analogue of the game-gradient field is the
\emph{payoff-difference field}, which records the directions of local payoff
improvement across signals and actions. Unlike in normal-form games, this field
depends on the Receiver's posterior belief, which is determined by Bayes' rule
from the Sender's strategy. Consequently, the payoff-difference field is not
multilinear, is generally not affine, and is only canonically defined on the
region where every signal is reached with positive probability. We call this
region the Bayes-defined region.\footnote{Outside this set, the Receiver's beliefs
are not pinned down by Bayes' rule; which beliefs should be considered rational is
a central issue in the signaling-games literature
\citep{cho1987,sobel2020signaling}.}

This change in geometry leads to a convergence puzzle. In our experiments,
standard online learning algorithms converge reliably to strict perfect Bayesian
equilibria ($\PBE$), but fail to converge to non-strict ones
(Section~\ref{sec:experiments}). Strict $\PBE$ are pure-strategy equilibria in which
each Sender type chooses a unique signal and the Receiver has a unique best
response at every signal. They include important separating and pooling
equilibria, such as the Riley outcome in Spence's job-market signaling model
\citep{spence1978job,riley1979}. However, this observed convergence cannot be
explained by the global geometric conditions that are standard tools for analyzing
first-order learning in normal-form games, as these do not extend directly to
signaling games.
Indeed, Theorem~\ref{thm:negative-K2} shows that even with two types, two
signals, and two actions, there are signaling games and strict $\PBE$ for which
neither monotonicity nor the Minty variational inequality holds on the
Bayes-defined region. Understanding why learning nevertheless converges
near strict $\PBE$ is the motivating question of this paper.

\subsection{Our Results}\label{subsec:Results}

Our first result shows that the standard global geometry used to prove
last-iterate convergence in normal-form games does not extend to signaling
games: even in binary signaling games, monotonicity and the Minty variational
inequality can fail at a strict $\PBE$ (Theorem~\ref{thm:negative-K2}). We therefore
turn to local geometry.
Our main
result shows that, in finite signaling games, a separating $\PBE$ is variationally stable if
and only if it is strict
(Theorem~\ref{thm:StrictPBE_equiv_VariationStable}). This parallels the
characterization of variationally stable strategy profiles as strict Nash
equilibria in finite normal-form games by
\citet[Proposition~5.2]{mertikopoulos2019learning}, but the extension is not
formal: in signaling games the payoff-difference field depends on Bayes-updated
posterior beliefs, is not multilinear, and is only canonically defined on the
Bayes-defined region. Thus, our result shows that strictness remains the
right local geometric condition for first-order learning even in this Bayesian
extensive-form setting. The proof is quantitative in the strict-$\PBE$-to-stability
direction: we derive an explicit radius of the stability neighborhood in terms of
primitive game parameters and the payoff-difference field
(Theorem~\ref{thm:StictPBE_implies_VS}). Combining this characterization with
the convergence result of \citet[Theorem~4.10]{mertikopoulos2019learning} yields
local last-iterate convergence of regularized dual averaging and online mirror
descent \citep{zhou2017mirror} to every strict separating $\PBE$ from sufficiently close
initializations.

In Section \ref{sec:convergence}, we move beyond local convergence guarantees. We identify a class of signaling games (which includes the example from Theorem~\ref{thm:negative-K2} that fails the Minty condition) for which, as we show in Theorem \ref{thm:unique-EFCCE}: the unique $\PBE$ is also the unique extensive-form coarse correlated equilibrium ($\EFBCCE$) \citep{FarinaBS20}. Since the broader $\EFBCCE$ polytope admits average-iterate convergence guarantees under no-coarse-trigger-regret dynamics~\citep{AnagnostidesFarinaKroerCelliSandholm2022}, and since under our assumptions this polytope collapses to the unique $\PBE$, we obtain a \emph{global} average-iterate convergence guarantee to the $\PBE$.

Finally, our experiments on canonical binary signaling games compare projected gradient ascent, online mirror ascent, optimistic mirror ascent, and PPO. The results support the theory: strict pure separating $\PBE$ are reliable attractors of standard first-order learning dynamics, whereas convergence to non-strict equilibria is fragile and algorithm-dependent. Interestingly, convergence is also observed in families with strict pooling or multiple pure PE, both of which lie outside the scope of Theorem \ref{thm:StrictPBE_equiv_VariationStable}. We also observe that optimistic (entropic) mirror descent can converge to a mixed equilibrium. We emphasize that the convergence for projected gradient ascent and PPO is only observed empirically.

Table~\ref{tab:game-classes-convergence} contains a summary of the results.

\begin{table}[ht!]
\centering
\caption{Overview of theoretical and empirical convergence results for the signaling-game classes considered in this paper. \textbf{Theoretical (Sections \ref{sec:last-iterate} and \ref{sec:convergence}):} $X^\circ(\mathcal{G})$ is the Bayes-defined region of the game \eqref{eq:Xcirc} and $q^{\star}$ is a $\PBE$.
\textbf{Empirical (Section~\ref{sec:experiments}):} $\checkmark$ = reliably converges, $\circ$ = initialization-dependent / partial, $\times$ = does not converge. The Minty condition (condition (ii) in Definition \ref{def:variational-conditions}) fails for a class, if there is at least one instance of that class that does not satisfy it.}
\label{tab:game-classes-convergence}
\small
\setlength{\tabcolsep}{4pt}
\renewcommand{\arraystretch}{1.15}
\begin{tabular}{p{4.0cm}p{3.3cm}p{3.0cm}p{1.8cm}p{2.0cm}}
\toprule
\textbf{Class of signaling games} & \textbf{Structural condition} & \textbf{Theoretical} & \textbf{Empirical} & \textbf{Result} \\
\midrule
unique separating
& strict $\PBE$ \newline ($q^\star \in X^\circ(\mathcal{G})$)
& Minty fails  \newline Variational Stability
& PGA $\checkmark$ \newline OMA $\checkmark$ \newline Opt.\,OMA $\checkmark$ \newline PPO $\checkmark$
& Thms.~\ref{thm:negative-K2},~\ref{thm:StrictPBE_equiv_VariationStable}; \newline experimental \\
\midrule
multiple separating; \newline
pooling-\&-separating
& strict $\PBE$ \newline ($\exists q^\star \in X^\circ(\mathcal{G})$)
& Unknown
& PGA $\checkmark$ \newline OMA $\checkmark$ \newline Opt.\,OMA $\checkmark$ \newline PPO $\checkmark$
& experimental \\
\midrule
unique pooling
& strict $\PBE$ \newline ($q^\star \notin X^\circ(\mathcal{G})$)
& Unknown
& PGA $\checkmark$ \newline OMA $\checkmark$ \newline Opt.\,OMA $\checkmark$ \newline PPO $\checkmark$
& experimental \\
\midrule
multiple pooling
& strict $\PBE$ \newline ($q^\star \notin X^\circ(\mathcal{G})$)
& Unknown
& PGA $\circ$ \newline OMA $\circ$ \newline Opt.\,OMA $\circ$ \newline PPO $\checkmark$
& experimental \\
\midrule
honest-signaling \newline
(Assumptions~\ref{assum:receiver-unique-best-response} and \ref{assum:sender-dominance})
& unique strict $\PBE$ \newline ($q^\star \in X^\circ(\mathcal{G})$)
&  Minty fails \newline Variational Stability \newline Unique EFCCE
& PGA $\checkmark$ \newline OMA $\checkmark$ \newline Opt.\,OMA $\checkmark$ \newline PPO $\checkmark$
& Thms.~\ref{thm:negative-K2},~\ref{thm:StrictPBE_equiv_VariationStable},~\ref{thm:unique-EFCCE}; \newline experimental \\
\midrule
unique semi-separating
& mixed $\PBE$
& Unknown
& PGA $\times$ \newline OMA $\times$ \newline Opt.\,OMA $\circ$ \newline PPO $\times$
& experimental \\
\bottomrule
\end{tabular}
\end{table}

\subsection{Related Work}
Our paper connects three literatures: learning in extensive-form games, convergence in monotone\slash variational-inequality game settings and signaling games as a minimal model of strategic interaction under asymmetric information. We cover the essential references on variational stability in Section \ref{subsec:limitations}.

\paragraph{Learning in extensive-form games.}
The dominant algorithmic approach to learning in extensive-form games is based
on regret minimization. In two-player zero-sum games, variants of
counterfactual regret minimization (CFR) exploit the recursive structure of the
game tree and yield average-iterate convergence to Nash equilibrium
\citep{ZinkevichJohansonBowlingPiccione2007}. In general-sum extensive-form
games, however, uncoupled no-regret guarantees typically lead not to Nash equilibrium but
to extensive-form correlated solution concepts. In particular,
trigger-regret minimization yields convergence of empirical play to
extensive-form correlated equilibrium (EFCE), while coarse-trigger-regret
minimization over sequence-form strategy spaces yields extensive-form coarse
correlated equilibrium (EFCCE) \citep{vonstengel08, FarinaBS20,
farina2022simple_jacm, FarinaCelliGattiSandholm2021}. Finally, to our
knowledge, last-iterate convergence to equilibrium in extensive-form games has
been studied only in the zero-sum case \citep{liu2023the, ren2025last,
mengefficient}.

\paragraph{Variational inequalities and monotone game learning.}
A large literature explains convergence of first-order methods in normal-form
games using structural conditions such as monotonicity and the Minty
variational inequality. These conditions play a role analogous to convexity in
optimization and underlie convergence guarantees for projection methods,
extragradient methods, mirror descent, and optimistic variants
\citep{bauschkeConvexAnalysisMonotone2017, tseng1995linear, adil2022optimal,
lin2022perseus, ratliff2013characterization, chasnov2019convergence,
strodiot2016class, Song2020OptDualExtrapolation, ye2022infeasible}.
If a game is strongly monotone, then simple first-order algorithms such as
OMA and PGA converge. If a game satisfies the Minty condition globally,
extragradient algorithms are known to converge. These are
the main known sufficient conditions for obtaining equilibrium
convergence.
Our paper analyzes monotonicity and the Minty condition in extensive-form
signaling games and draws on techniques from dynamical systems to certify
last-iterate convergence locally.

\paragraph{Signaling games and learning under asymmetric information.}
Signaling games are a canonical model of strategic interaction under asymmetric
information, with applications throughout economics and related fields
\citep{cho1987, spence1978job, mailath1987incentive, sobel2020signaling}. They are also a
natural minimal testbed for our question: they are simple enough to analyze
explicitly, yet already contain the key ingredients absent from normal-form
models, namely sequential play, belief updating, and Bayes-consistency
constraints. A large economics literature studies equilibrium structure and
refinements in signaling games, including perfect Bayesian equilibrium,
sequential equilibrium, the Intuitive Criterion, and D1
\citep{cho1987, cho1990, ramey1996}. Our focus is different. We do not study
equilibrium refinement or selection through off-path beliefs per se; instead,
we use signaling games as the simplest incomplete-information extensive-form
environment in which Bayesian updating changes the learning geometry. There is also prior work on learning in signaling and sender--receiver games.
Some of this literature studies how learning shapes off-path beliefs and
equilibrium selection in signaling environments
\citep{FudenbergHe2018, FudenbergHe2020}, while older work emphasizes that
uncoupled learning in extensive-form games may converge only to
self-confirming equilibrium rather than Nash equilibrium
\citep{FudenbergLevine1993, FudenbergKreps1995, HendonJacobsenSloth1996}.

\section{Preliminaries}\label{sec:model}

\paragraph{Finite signaling games} A signaling game consists of a Sender and a Receiver. Nature first draws the
Sender's type \(\theta \in \Theta\) from a commonly known prior
\(\pi \in \Delta(\Theta)\)\footnote{For a finite set $X$, $\Delta(X)$ denotes the simplex of probability distributions over $X$.} with $\pi(\theta)>0$ for all $\theta \in \Theta$.  The Sender observes \(\theta\) and chooses a signal \(s \in \Sigma\). The Receiver observes \(s\),
and then chooses an action \(a \in A\). Utilities are given by
functions $u_S,u_R : \Theta \times \Sigma \times A \to \mathbb{R}$. We focus on the finite case, where
\(\Theta\), \(\Sigma\), and \(A\) are finite sets and we assume $|\Theta|=|\Sigma|=|A| = K$ for a $K \in \mathbb{N}_{\geq 2}$. Explicitly, we denote the sets as $\Theta=\{\theta_1, \dots, \theta_K\}$, $\Sigma = \{s_1,\dots,s_K\}$, $A=\{a_1, \dots a_K\}$. We refer to such a signaling game by the tuple $\mathcal{G} = (\Theta, \Sigma, A, \pi, u_S, u_R)$.

A behavioral strategy of the Sender is a conditional probability distribution $q_S(\cdot \mid \theta) \in \Delta(\Sigma)$ for each $\theta \in \Theta$, and a behavioral strategy of the Receiver is a conditional distribution $q_R(\cdot \mid s) \in \Delta(A)$ for each $s \in \Sigma$. A \emph{strategy profile} is a pair $q = (q_S, q_R)$. Since $\mathcal{G}$ is finite, its strategy space is $X(\mathcal{G}) \;:=\; \prod_{\theta \in \Theta} \Delta(\Sigma) \;\times\; \prod_{s \in \Sigma} \Delta(A)$.

Given a Sender strategy $q_S$, signal $s$ is said to be \emph{on-path} under $q_S$ if it has positive marginal probability under the prior, $\sum_{\theta \in \Theta} \pi(\theta)\,q_S(s \mid \theta) > 0$. Whenever $s$ is on-path, the Receiver's posterior is determined by Bayes' rule:
\begin{equation} \label{eq. posterior update}
\mu(\theta \mid s)
\;=\;
\frac{\pi(\theta)\, q_S(s\mid \theta)}{\sum_{\theta' \in \Theta} \pi(\theta')\, q_S(s\mid \theta')}.
\end{equation}
Strategy profiles in $X(\mathcal{G})$ for which every signal is on-path under $q_S$ form the \emph{Bayes-defined region}:
\begin{equation}
\label{eq:Xcirc}
X^{\circ}(\mathcal{G}) \;:=\; \Bigl\{\, q \in X(\mathcal{G}) \;:\; \sum_{\theta \in \Theta} \pi(\theta)\,q_S(s \mid \theta) > 0 \text{ for all } s \in \Sigma \,\Bigr\}.
\end{equation}
\begin{restatable}{lemma}{xcircconvexopen}\label{lem:Xcirc-convex-open}
For every signaling game $\mathcal{G}$, $X^{\circ}(\mathcal{G})$ is convex and open in $X(\mathcal{G})$.
\end{restatable}
\begin{proof}
    Fix $q', q'' \in X^{\circ}(\mathcal{G})$ and $t \in [0, 1]$, and set $\tilde q = (1-t)\,q' + t\,q''$. For every $s \in \Sigma$, we have:
\[
\sum_{\theta \in \Theta} \pi(\theta)\,\tilde q_S(s \mid \theta) \;=\; (1-t)\sum_{\theta \in \Theta} \pi(\theta)\,q'_S(s \mid \theta) \;+\; t\sum_{\theta \in \Theta} \pi(\theta)\,q''_S(s \mid \theta) \;>\; 0.
\]
Here, the strict inequality follows from the fact that both $q',q'' \in X^{\circ}(\mathcal{G})$. Hence, $\tilde{q} \in  X^{\circ}(\mathcal{G})$ and convexity follows.

To complete the proof, we prove that $X^{\circ}(\mathcal{G})$ is open in $X(\mathcal{G})$. For each $s \in \Sigma$,  $\phi_s(q) := \sum_{\theta \in \Theta} \pi(\theta)\,q_S(s \mid \theta)$ is continuous on $X(\mathcal{G})$. Therefore, the set $\{q \in X(\mathcal{G}) : \phi_s(q) > 0\}$ is open in $X(\mathcal{G})$. As a finite intersection of these sets, $X^{\circ}(\mathcal{G})$ is open in $X(\mathcal{G})$.
\end{proof}
Lemma \ref{lem:Xcirc-convex-open} will be useful in the analysis that follows.
Finally, we write $\supp(q_S(\cdot \mid \theta)):=\{s \in \Sigma: q_S(s \mid \theta)> 0\}$ and $\supp(q_R(\cdot \mid s)) := \{a \in A: q_R(a \mid s) > 0\}$.

\paragraph{Solution concepts}

A \emph{belief system} for $\mathcal{G}$ is a map $\mu : \Sigma \to \Delta(\Theta)$ assigning to every signal $s \in \Sigma$ a posterior $\mu(\cdot \mid s) \in \Delta(\Theta)$ over Sender types. An \emph{assessment} of $\mathcal{G}$ is a pair $(q, \mu)$ consisting of a strategy profile $q = (q_S, q_R) \in X(\mathcal{G})$ and a belief system $\mu$ for $\mathcal{G}$.

\begin{definition}\label{def. Perfect Bayesian Equilibrium}
An assessment $(q, \mu)$ of $\mathcal{G}$ is a \emph{Perfect Bayesian Equilibrium} ($\PBE$) of $\mathcal{G}$ if:
\begin{enumerate}[label=(\roman*)]
\item For every $\theta \in \Theta$ and every $s \in \Sigma$,
\[
\begin{aligned}
\supp\bigl(q_S(\cdot \mid \theta)\bigr)
&\subseteq \operatorname*{arg\,max}_{s' \in \Sigma} \sum_{a \in A} u_S(\theta, s', a)\, q_R(a \mid s'), \\
\supp\bigl(q_R(\cdot \mid s)\bigr)
&\subseteq \operatorname*{arg\,max}_{a \in A} \sum_{\theta' \in \Theta} u_R(\theta', s, a)\, \mu(\theta' \mid s).
\end{aligned}
\]
\item At every signal $s \in \Sigma$ that is on-path under $q_S$, the belief $\mu(\cdot \mid s)$ is given by Bayes' rule~\eqref{eq. posterior update}. For the remaining signals
the belief $\mu(\cdot \mid s) \in \Delta(\Theta)$ is unrestricted.
\end{enumerate}
\end{definition}

Condition~(i) is known as \emph{sequential rationality} and condition~(ii) as \emph{consistency}. We use $\mu$ for both the Bayes posterior in $\eqref{eq. posterior update}$ and the belief system of an assessment: on $X^{\circ}(\mathcal{G})$ these coincide, since every signal is on-path, so the belief system $\mu$ is uniquely determined by Bayes' rule. Accordingly, we identify a $\PBE$ assessment $(q, \mu)$ on $X^{\circ}$ with its strategy profile $q$, and refer to $q^{\star} \in X^{\circ}(\mathcal{G})$ simply as a $\PBE$.
In the signaling game literature, a $\PBE$ is further classified as separating, pooling, or semi-separating according to whether the Sender's signal reveals her type.

\begin{definition}\label{def. signaling games PBE}
A $\PBE$ $(q, \mu)$ of $\mathcal{G}$ is:
\begin{enumerate}[label=(\roman*)]
\item \emph{Separating} if for every $\theta \in \Theta$ there exists a signal $s \in \Sigma$ such that
$
\supp\bigl(q_S(\cdot \mid \theta)\bigr) = \{s\}
\quad$ and $
s \notin \supp\bigl(q_S(\cdot \mid \theta')\bigr)$ for every $ \theta' \in \Theta \setminus \{\theta\}$.

\item \emph{Pooling} if there exists a signal $s \in \Sigma$ such that $\supp(q_S(\cdot \mid \theta)) = \{s\}$ for every $\theta \in \Theta$.

\item \emph{Semi-separating} if it is neither separating nor pooling.
\end{enumerate}
\end{definition}

In a separating $\PBE$, each type plays a pure signal that no other type sends, so observing the signal uniquely identifies the Sender's type. In a pooling $\PBE$, all types send the same signal, so observing it conveys no information about the type.

Up to relabeling, the \emph{Riley $\PBE$} is the assessment $(q^{\star}, \mu)$ in which type $\theta_k$ plays signal $s_k$ and the Receiver responds with $a_k$ for all $k \in \{1, \ldots, K\}$ i.e.,
    \begin{equation}\label{eq:riley}
            q_S^\star(s_k \mid \theta) = \mathbf{1}\{\theta = \theta_k\},
    \qquad
    q_R^\star(a \mid s_k) = \mathbf{1}\{a = a_k\},
    \qquad
    \mu^\star(\theta \mid s_k) = \mathbf{1}\{\theta = \theta_k\}.
    \end{equation}
We define the set of $\PBE$ strategy profiles of $\mathcal{G}$ as
\[
\SPBE \;=\; \bigl\{ q \in X(\mathcal{G}) \;:\; \exists \mu : \Sigma \to \Delta(\Theta) \text{ such that } (q,\mu) \text{ is a } \PBE \text{ of } \mathcal{G} \bigr\}.
\]
We say $\mathcal{G}$ has a \emph{unique $\PBE$} if $\SPBE$ is a singleton. We call the strategy profile $q^{\star}$ of the Riley $\PBE$ the \RO \citep{riley1979}. A $\PBE$ $q^{\star}$ is \emph{pure} if $|\supp q^{\star}_S(\cdot \mid \theta)| = |\supp q^{\star}_R(\cdot \mid s)| = 1$ for every $\theta \in \Theta$ and $s \in \Sigma$; write $\sigma^{\star}(\theta)$ and $\alpha^{\star}(s)$ for the unique elements of these supports. A pure $\PBE$ is \emph{strict} if $u_S(\theta, \sigma^{\star}(\theta), \alpha^{\star}(\sigma^{\star}(\theta))) > u_S(\theta, s, \alpha^{\star}(s))$ for every $\theta \in \Theta$ and $s \neq \sigma^{\star}(\theta)$, and $\sum_{\theta} \mu^{\star}(\theta \mid s)\, u_R(\theta, s, \alpha^{\star}(s)) > \sum_{\theta} \mu^{\star}(\theta \mid s)\, u_R(\theta, s, a)$ for every $s \in \Sigma$ and $a \neq \alpha^{\star}(s)$.

We next define the Bayesian extensive-form coarse correlated equilibrium (EFCCE), introduced by \citet{FarinaBS20}. Given $\mathcal{G}$, a \emph{recommendation device} is a probability distribution $\cor \in \Delta(\Sigma^\Theta \times A^\Sigma)$ over pairs of pure strategies $(\sigma, \alpha)$, with $\sigma : \Theta \to \Sigma$ and $\alpha : \Sigma \to A$. In an EFCCE, a pair $(\sigma, \alpha) \sim \cor$ is drawn, and each player commits to follow the recommendation when it is their turn, without observing it.

\begin{definition}[\citealp{FarinaBS20}]\label{def:efcce}
A probability distribution $\cor \in \Delta(\Sigma^\Theta \times A^\Sigma)$ is a
Bayesian extensive-form coarse correlated equilibrium ($\EFBCCE$) if:
\begin{enumerate}
    \item For every type $\theta \in \Theta$ and every signal $s' \in \Sigma$:
    \begin{equation}\label{eq:efcce-sender}
    \mathbb{E}_{(\sigma,\alpha) \sim  \cor}\bigl[
      u_S\bigl(\theta,\, \sigma(\theta),\, \alpha(\sigma(\theta))\bigr)
    \bigr]
    \;\ge\;
    \mathbb{E}_{(\sigma,\alpha) \sim \cor}\bigl[
      u_S\bigl(\theta,\, s',\, \alpha(s')\bigr)
    \bigr].
    \end{equation}

    \item For every signal $s \in \Sigma$ with $\sum_{\theta \in \Theta}\pi(\theta)\Pr_{(\sigma, \alpha) \sim Q}[\sigma(\theta)=s] > 0$ and every action $a' \in A$:
    \begin{equation}\label{eq:efcce-receiver}
    \sum_{\theta \in \Theta}
    \mu(\theta \mid s)\,
    \mathbb{E}_{(\sigma,\alpha) \sim \cor}\bigl[
      u_R(\theta,\, s,\, \alpha(s)) \mid \sigma(\theta)=s
    \bigr]
    \;\ge\;
    \sum_{\theta \in \Theta}
    \mu(\theta \mid s)\,
    u_R(\theta, s, a'),
    \end{equation}
    with $\mu(\theta | s)$ defined according to \eqref{eq. posterior update}\footnote{For a recommendation device $\cor$, the marginal $q_S(s \mid \theta)$ in~\eqref{eq. posterior update} is replaced by the projection $\Pr_{(\sigma, \alpha) \sim \cor}[\sigma(\theta) = s]$.}.
\end{enumerate}
\end{definition}
The conditions in Definition \ref{def:efcce} require that no Sender profitably deviates to a fixed $s' \in \Sigma$ \eqref{eq:efcce-sender}, and that no Receiver, conditional on the recommended signal being $s \in \Sigma$, profitably deviates to a fixed  action $a' \in A$ \eqref{eq:efcce-receiver}.%

Every $\PBE$ strategy $q^{\star}$ also embeds into the EFCCE polytope as the product recommendation device with $\sigma(\theta) \sim q^{\star}_S(\cdot \mid \theta)$ and $\alpha(s) \sim q^{\star}_{R}(\cdot \mid s)$, with all draws independent. Then, the EFCCE conditions \eqref{eq:efcce-sender}-\eqref{eq:efcce-receiver} reduce to the on-path\footnote{Condition~\eqref{eq:efcce-receiver} only constrains on-path
signals; off-path receiver behavior is unrestricted, in contrast to
perfect correlated solution concepts~\citep{marchesi21}.} $\PBE$ sequential rationality conditions in Definition~\ref{def. Perfect Bayesian Equilibrium}. Therefore, $\SPBE \subseteq \SEFBCCE$.
\paragraph{Payoff-difference field}

To study first-order learning dynamics via geometric conditions, we work with the induced \emph{payoff-difference field}.

\begin{definition}\label{def:PayoffDifferenceField}
Given a \emph{reference profile} $(\bar{\sigma}, \bar{\alpha})$ with $\bar{\sigma}: \Theta \to \Sigma$ and $\bar{\alpha}: \Sigma \to A$, the \emph{payoff-difference field} of $\mathcal{G}$ relative to $(\bar{\sigma}, \bar{\alpha})$ is the map $g : X^{\circ}(\mathcal{G}) \to \mathbb{R}^{\Theta \times \Sigma} \times \mathbb{R}^{\Sigma \times A}$ with components, for $\theta \in \Theta$, $s \in \Sigma$, $a \in A$,
\begin{align}
g_S(\theta, s;\, q) &\;:=\; \sum_{a' \in A} u_S(\theta, s, a')\,q_R(a' \mid s) \;-\; \sum_{a' \in A} u_S(\theta, \bar{\sigma}(\theta), a')\,q_R\bigl(a' \mid \bar{\sigma}(\theta)\bigr), \label{eq:g-sender} \\
g_R(s, a;\, q) &\;:=\; \sum_{\theta \in \Theta} \bigl[u_R(\theta, s, a) - u_R\bigl(\theta, s, \bar{\alpha}(s)\bigr)\bigr]\,\mu(\theta \mid s).\label{eq:g-receiver}
\end{align}
\end{definition}

In words, $g_S(\theta, s; q)$ is the Sender's expected payoff at type $\theta$ from signal $s$ minus that from the reference signal $\bar{\sigma}(\theta)$, and $g_R(s, a; q)$ is the Receiver's posterior expected payoff at signal $s$ from action $a$ minus that from the reference action $\bar{\alpha}(s)$. Subtracting the reference payoff shifts all coordinates of $g_S(\theta, \cdot\,; q)$ (resp.\ $g_R(s, \cdot\,; q)$) by the same constant, which changes neither best responses nor any of the variational conditions used in Section~\ref{sec:last-iterate} (see Proposition~\ref{lem:reference-invariance}). Up to this shift and a positive rescaling (by $\pi(\theta)$ for the Sender, and by the probability $\sum_{\theta \in \Theta} \pi(\theta)\, q_S(s \mid \theta)$ that $s$ is reached for the Receiver), $g$ is the gradient of each player's expected utility, i.e., the input to the first-order dynamics we study.

The key difference from the normal-form setting (see e.g., \citet{mertikopoulos2019learning}) is that $g$ depends on posterior beliefs induced by Bayes' rule. As a result, $g$ need not be multilinear, need not be globally defined on all of $X(\mathcal{G})$, and can have geometric properties that differ from those in normal-form games. Nevertheless, as we show in Lemma \ref{lem:g-continuous}, $g$ retains continuity on $X^{\circ}(\mathcal{G})$. This is the basic regularity needed for our analysis in Section \ref{sec:last-iterate}.

\begin{restatable}{lemma}{gcontinuous}\label{lem:g-continuous}
The payoff-difference field $g$ is continuous on $X^{\circ}(\mathcal{G})$.
\end{restatable}
\begin{proof}
For every $\theta \in \Theta$ and every $s \in \Sigma$, $g_S(\theta, s;\, q)$ (defined in~\eqref{eq:g-sender}) is linear in $q_R$ and does not depend on $q_S$. Linear functions are continuous, so $g_S(\theta, s;\, \cdot)$ is continuous on $X^{\circ}(\mathcal{G})$.

For every $s \in \Sigma$ and every $a \in A$, $g_R(s, a;\, q)$ (defined in~\eqref{eq:g-receiver}) depends on $q$ only through the Bayes posterior
\[
\mu(\theta \mid s;\, q) \;=\; \frac{\pi(\theta)\,q_S(s \mid \theta)}{\sum_{\theta' \in \Theta} \pi(\theta')\,q_S(s \mid \theta')}.
\]
The numerator and denominator are linear in $q_S$, and the denominator is strictly positive on $X^{\circ}(\mathcal{G})$ by definition. Therefore, $\mu(\theta \mid s;\, q)$ is continuous in $q$ on $X^{\circ}(\mathcal{G})$, and so is $g_R(s, a;\, q)$.

Since both components are continuous, $g$ is continuous on $X^{\circ}(\mathcal{G})$.
\end{proof}

The payoff-difference field of Definition~\ref{def:PayoffDifferenceField} depends on the reference profile $(\bar{\sigma}, \bar{\alpha})$ only conventionally: every quantity used in the analysis of the paper does not depend on this choice. To this end, we show the following proposition; its proof can be found in Appendix~\ref{app:reference-invariance}.

\begin{restatable}{proposition}{referenceinvariance}\label{lem:reference-invariance}
Let $(\bar{\sigma}_1, \bar{\alpha}_1)$ and $(\bar{\sigma}_2, \bar{\alpha}_2)$ be two reference profiles, and let $g^{(1)}, g^{(2)}$ denote their payoff-difference fields. Then for every $q, q' \in X^{\circ}(\mathcal{G})$,
$
\langle g^{(1)}(q'),\, q - q'\rangle \;=\; \langle g^{(2)}(q'),\, q - q'\rangle.
$
\end{restatable}

\section{Last-iterate Convergence}\label{sec:last-iterate}%

In this section, we study last-iterate convergence of first-order learning dynamics in signaling games. A classical perspective, which we also adopt, is to establish geometric conditions on $g$ (Definition~\ref{def:PayoffDifferenceField}).

\subsection{Limitations to Global Last-iterate Convergence}\label{subsec:limitations}

In normal-form games, the standard route to global last-iterate convergence of first-order learning dynamics is to verify a global geometric condition on the payoff-difference field. In Definition~\ref{def:variational-conditions}, we highlight the two such conditions central to the literature; throughout, $\langle \cdot, \cdot\rangle$ denotes the standard Euclidean inner product on $\mathbb{R}^{\Theta \times \Sigma} \times \mathbb{R}^{\Sigma \times A}$.

\begin{definition}\label{def:variational-conditions}
Let $\mathcal{G}$ be a signaling game and let $g$ be its payoff-difference field.
\begin{enumerate}[label=(\roman*)]
\item $g$ is \emph{monotone} if $\langle g(q) - g(q'),\, q - q'\rangle \le 0$ for all $q, q' \in X^{\circ}(\mathcal{G})$.
\item $q^{\star} \in X^{\circ}(\mathcal{G})$ satisfies the \emph{Minty condition} if $\langle g(q),\, q - q^{\star}\rangle \le 0$ for every $q \in X^{\circ}(\mathcal{G})$.
\end{enumerate}
\end{definition}

Standard theory on variational inequalities (see e.g.,~\citep[Chapter 2]{facchinei2003finite}) relates the two: if $g$ is monotone and $q^{\star}$ is a $\PBE$ of $\mathcal{G}$, then $q^{\star}$ satisfies the Minty condition; conversely, the Minty condition at $q^{\star}$ implies that $q^{\star}$ is a $\PBE$ of $\mathcal{G}$.

In normal-form games, monotonicity yields last-iterate convergence of dual averaging and mirror descent~\citep{bervoets2020learning, hofbauer2009stable, mertikopoulos2019learning}, while the Minty condition yields last-iterate convergence of extragradient and optimistic methods~\citep{daskalakis2018training, mokhtari2020unified}. Somewhat surprisingly, we show in Theorem \ref{thm:negative-K2} that for finite signaling games, even for the simplest case of $|\Theta| = |\Sigma| = |A| = 2$ there are instances where these conditions fail.

\begin{restatable}{theorem}{negativektwo}\label{thm:negative-K2}
There exists a signaling game $\mathcal{G}$ with $|\Theta| = |\Sigma| = |A| = 2$ and a strict separating $\PBE$ $q^{\star} \in X^{\circ}(\mathcal{G})$ at which the Minty condition fails.
\end{restatable}

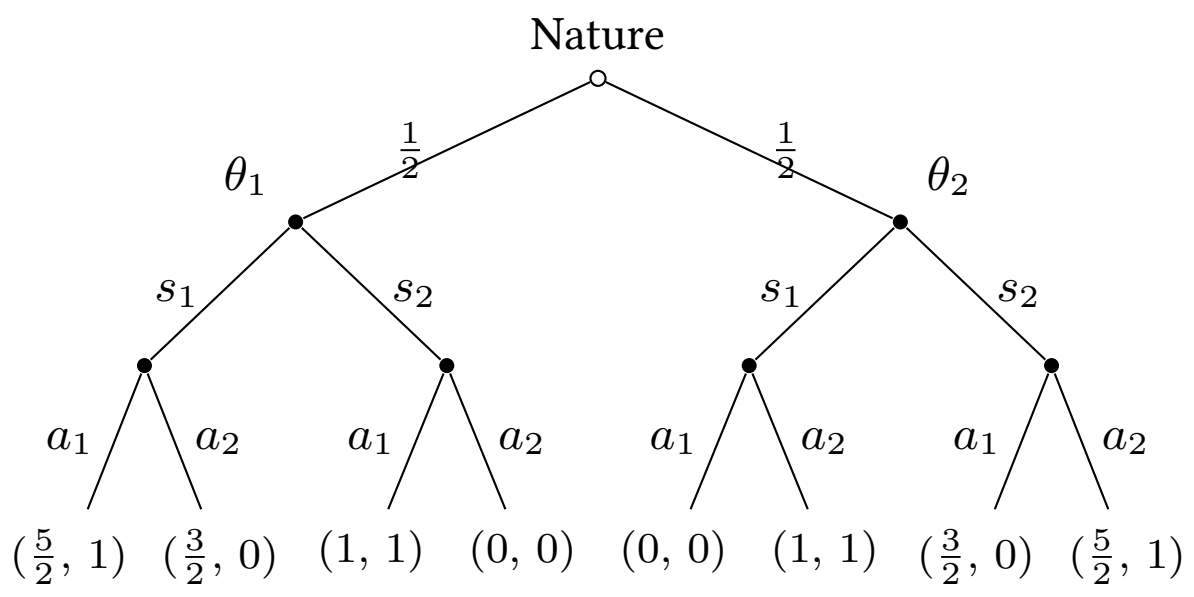
\begin{figure}[t]
\centering
\begin{tikzpicture}[
    font=\footnotesize,
    level 1/.style={sibling distance=4cm, level distance=0.95cm},
    level 2/.style={sibling distance=2cm, level distance=0.95cm},
    level 3/.style={sibling distance=1cm, level distance=0.95cm},
    hollow/.style={circle, draw, inner sep=1pt},
    solid/.style={circle, fill=black, inner sep=1pt},
]
\node[hollow, label=above:{Nature}] {}
  child {node[solid, label=above left:{$\theta_1$}] {}
    child {node[solid] {}
      child {node[below, font=\scriptsize] {$(\tfrac{5}{2},\,1)$} edge from parent node[left] {$a_1$}}
      child {node[below, font=\scriptsize] {$(\tfrac{3}{2},\,0)$} edge from parent node[right] {$a_2$}}
      edge from parent node[left] {$s_1$}
    }
    child {node[solid] {}
      child {node[below, font=\scriptsize] {$(1,\,1)$} edge from parent node[left] {$a_1$}}
      child {node[below, font=\scriptsize] {$(0,\,0)$} edge from parent node[right] {$a_2$}}
      edge from parent node[right] {$s_2$}
    }
    edge from parent node[left] {$\tfrac{1}{2}$}
  }
  child {node[solid, label=above right:{$\theta_2$}] {}
    child {node[solid] {}
      child {node[below, font=\scriptsize] {$(0,\,0)$} edge from parent node[left] {$a_1$}}
      child {node[below, font=\scriptsize] {$(1,\,1)$} edge from parent node[right] {$a_2$}}
      edge from parent node[left] {$s_1$}
    }
    child {node[solid] {}
      child {node[below, font=\scriptsize] {$(\tfrac{3}{2},\,0)$} edge from parent node[left] {$a_1$}}
      child {node[below, font=\scriptsize] {$(\tfrac{5}{2},\,1)$} edge from parent node[right] {$a_2$}}
      edge from parent node[right] {$s_2$}
    }
    edge from parent node[right] {$\tfrac{1}{2}$}
  };
\end{tikzpicture}
\caption{Game tree for the instance used in Theorem~\ref{thm:negative-K2}. Terminal payoffs are $(u_S, u_R)$.}
\label{fig:negative-K2-tree}
\end{figure}

The instance  we use in Theorem \ref{thm:negative-K2} is the following (see Figure~\ref{fig:negative-K2-tree})%
: the binary signaling game with prior $\pi(\theta_1) = \pi(\theta_2) = 1/2$ and utilities $u_R(\theta_k, s, a_j) = \mathbf{1}\{j = k\}$ and $u_S(\theta_k, s_i, a_j) = \tfrac{3}{2}\,\mathbf{1}\{i = k\} + \mathbf{1}\{j = k\}$. It is a remarkably simple instance: the Riley outcome $q^{\star}$ in \eqref{eq:riley} is its unique strict $\PBE$ strategy and the Sender has a dominant strategy to signal her type. Yet, the Minty condition (Definition~\ref{def:variational-conditions}(ii)) fails on $q^{\star}$. Since monotonicity of $g$ on $X^{\circ}(\mathcal{G})$ (Definition \ref{def:variational-conditions}(i)) implies Minty at every $\PBE$, Theorem \ref{thm:negative-K2} also rules out monotonicity of g on $X^{\circ}$.

\begin{proof}[Proof of Theorem~\ref{thm:negative-K2}]
Consider the signaling game $\mathcal{G}$ with $\Theta = \{\theta_1, \theta_2\}$, $\Sigma = \{s_1, s_2\}$, $A = \{a_1, a_2\}$, prior $\pi(\theta_1) = \pi(\theta_2) = 1/2$, and utilities
\[
u_R(\theta_k, s, a_j) \;=\; \mathbf{1}\{j = k\}, \qquad u_S(\theta_k, s_i, a_j) \;=\; \tfrac{3}{2}\,\mathbf{1}\{i = k\} + \mathbf{1}\{j = k\}.
\]
Let $q^{\star}$ be the Riley outcome, with $q_S^{\star}(s_k \mid \theta_k) = 1$ and $q_R^{\star}(a_k \mid s_k) = 1$, and let $\mu^{\star}$ be such that $\mu^{\star}(\theta_k \mid s_k) = 1$. Observe that $q^{\star} \in X^{\circ}$ since $\sum_{\theta \in \Theta}\pi(\theta)q^{\star}_S(s_k \mid  \theta)=\pi(\theta_k)=1/2>0$ holds for $k \in \{1,2\}$.

To verify that $q^{\star}$ is a $\PBE$ (Definition~\ref{def. Perfect Bayesian Equilibrium}), we check sequential rationality and consistency. Fix $k \in \{1, 2\}$. For every $i \in \{1, 2\}$, the Sender's expected utility at type $\theta_k$ from signal $s_i$ against $q_R^{\star}$ is $\sum_a u_S(\theta_k, s_i, a)\,q_R^{\star}(a \mid s_i) = u_S(\theta_k, s_i, a_i) = \tfrac{3}{2}\mathbf{1}\{i = k\} + \mathbf{1}\{i = k\} = \tfrac{5}{2}\mathbf{1}\{i = k\}$, uniquely maximized at $i = k$. For every $j \in \{1, 2\}$, the Receiver's expected utility at signal $s_k$ from action $a_j$ under belief $\mu^{\star}(\theta_k \mid s_k) = 1$ is $u_R(\theta_k, s_k, a_j) = \mathbf{1}\{j = k\}$, uniquely maximized at $j = k$. Consistency is immediate, since $\mu^{\star}$ equals the Bayes posterior at every on-path signal. Hence $q^{\star}$ is a strict $\PBE$.

To show that the Minty condition fails on $X^{\circ}(\mathcal{G})$, consider the profile $\bar q$ defined by $\bar q_S(s_j \mid \theta_k) = \mathbf{1}\{j \neq k\}$ and $\bar q_R(a_j \mid s_k) = \mathbf{1}\{j \neq k\}$ for every $j, k \in \{1, 2\}$. The signal marginals at $\bar q_S$ satisfy $\sum_{\theta \in \Theta} \pi(\theta)\,\bar q_S(s_k \mid \theta) = 1/2$ for every $k \in \{1, 2\}$, so $\bar q \in X^{\circ}(\mathcal{G})$, and Bayes' rule gives $\mu(\theta_2 \mid s_1;\, \bar q) = \mu(\theta_1 \mid s_2;\, \bar q) = 1$. With reference profile $(\bar\sigma, \bar\alpha) = (\sigma^{\star}, \alpha^{\star})$ (observe that this is without loss of generality by Proposition \ref{lem:reference-invariance}), we compute $g_S(\theta_k, s_j;\, \bar q)$ and $g_R(s_k, a_j;\, \bar q)$ for every $j, k \in \{1, 2\}$ with $j \neq k$. For the Sender,
\[
g_S(\theta_1, s_2;\, \bar q) \;=\; u_S(\theta_1, s_2, a_1) - u_S(\theta_1, s_1, a_2) \;=\; 1 - \tfrac{3}{2} \;=\; -\tfrac{1}{2},
\]
using $\bar q_R(a_1 \mid s_2) = 1$ and $\bar q_R(a_2 \mid s_1) = 1$, and by symmetry $g_S(\theta_2, s_1;\, \bar q) = -1/2$. For the Receiver,
\[
g_R(s_1, a_2;\, \bar q) \;=\; u_R(\theta_2, s_1, a_2) - u_R(\theta_2, s_1, a_1) \;=\; 1,
\]
using $\mu(\theta_2 \mid s_1;\, \bar q) = 1$, and by symmetry $g_R(s_2, a_1;\, \bar q) = 1$. By our choice of reference profile, $g_S(\theta_k, s_k;\, \bar q) = 0$ and $g_R(s_k, a_k;\, \bar q) = 0$ for every $k \in \{1, 2\}$, so these terms vanish in the inner product. The remaining four terms satisfy $\bar q_S(s_j \mid \theta_k) - q^{\star}_S(s_j \mid \theta_k) = 1$ and $\bar q_R(a_j \mid s_k) - q^{\star}_R(a_j \mid s_k) = 1$ for every $j \neq k$. Therefore,
\begin{align*}
  \langle g(\bar q),\, \bar q - q^{\star}\rangle \;&=\; g_S(\theta_1, s_2;\, \bar q) + g_S(\theta_2, s_1;\, \bar q) + g_R(s_1, a_2;\, \bar q) + g_R(s_2, a_1;\, \bar q)\\
  &\;=\; -\tfrac{1}{2} - \tfrac{1}{2} + 1 + 1 \;=\; 1 \;>\; 0,
\end{align*}
which violates the Minty condition at $q^{\star}$ (Definition~\ref{def:variational-conditions}(ii)).
\end{proof}

\subsection{Variational Stability}\label{subsec:V_S}

Theorem~\ref{thm:negative-K2} rules out the global geometric conditions typically used to obtain last-iterate convergence. Therefore, we turn to a \emph{local} condition called \emph{variational stability}~\citep{mertikopoulos2019learning}.

\begin{definition}[\citealp{mertikopoulos2019learning}]\label{def:VS}
A strategy profile $q^{\star} \in X^{\circ}(\mathcal{G})$ is variationally stable if there exists an open neighborhood $H \subseteq X^{\circ}(\mathcal{G})$ of $q^{\star}$ such that
$
\langle g(q),\, q - q^{\star}\rangle \le 0$ for every $q \in H,$ with strict inequality for $ q \in H \setminus \{q^{\star}\}$.

\end{definition}

For finite signaling games, the notion is well-posed by Lemma~\ref{lem:Xcirc-convex-open}: openness of $X^{\circ}(\mathcal{G})$ guarantees that every $q^{\star} \in X^{\circ}(\mathcal{G})$ has a non-trivial open neighborhood in $X^{\circ}(\mathcal{G})$.

We stress that variational stability and the Minty condition of
Definition~\ref{def:variational-conditions}(ii) are not directly comparable: the former
imposes a \emph{strict} inequality on a neighborhood of $q^{\star}$, while the
latter imposes a non-strict inequality on $X^{\circ}(\mathcal{G})$.
The two conditions also yield different algorithmic guarantees. Under
monotonicity or Minty, mirror descent and dual averaging can cycle in last
iterate (as in zero-sum bilinear games~\citep{mertikopoulos2018cycles, daskalakis2018training}),
while extragradient and optimistic-gradient methods recover last-iterate
convergence in these settings~\citep{mokhtari2020unified, daskalakis2018training}.
By contrast, under variational stability, FTRL dynamics \citep{mertikopoulos2019learning}  (in particular,
dual averaging with a strongly convex regularizer) and online mirror descent \citep{zhou2017mirror} attain last-iterate
convergence to $q^{\star}$ from initializations sufficiently close to
$q^{\star}$; see \citep[Section 4]{mertikopoulos2019learning} for a detailed
discussion.

For finite normal-form games, \citet[Proposition 5.2]{mertikopoulos2019learning} show that  variationally stable strategy profiles coincide with strict Nash equilibria. In our main result, Theorem~\ref{thm:StrictPBE_equiv_VariationStable}, we extend this characterization to strict $\PBE$ for finite signaling games.%

\begin{theorem}\label{thm:StrictPBE_equiv_VariationStable}
Let $\mathcal{G}$ be a signaling game and let $q^{\star} \in X^{\circ}(\mathcal{G})$. Then, $q^{\star}$ is variationally stable if and only if it is a strict $\PBE$.
\end{theorem}

We devote the rest of this section to proving Theorem~\ref{thm:StrictPBE_equiv_VariationStable}, showing first that every strict $\PBE$ $q^{\star} \in X^{\circ}(\mathcal{G})$ is variationally stable.

Fix a strict $\PBE$ $q^{\star} \in X^{\circ}(\mathcal{G})$. Since $|\supp q^{\star}_S(\cdot \mid \theta)| = |\supp q^{\star}_R(\cdot \mid s)| = 1$ for every $\theta \in \Theta$ and $s \in \Sigma$ (by definition of strict $\PBE$), we write $\sigma^{\star}(\theta)$ for the unique signal in $\supp q^{\star}_S(\cdot \mid \theta)$ and $\alpha^{\star}(s)$ for the unique action in $\supp q^{\star}_R(\cdot \mid s)$. Equivalently, this defines maps $\sigma^{\star}\colon \Theta \to \Sigma$ and $\alpha^{\star}\colon \Sigma \to A$ that capture $q^{\star}$. Throughout this section, we set $\bar{\sigma} = \sigma^{\star}$ and $\bar{\alpha} = \alpha^{\star}$.%

We now introduce auxiliary notation. Let
\begin{align*}
\delta(\mathcal{G}, q^{\star}) &:= \min\left\{\min_{\theta \in \Theta,\, s \neq \sigma^{\star}(\theta)} |g_S(\theta, s; q^{\star})|,\, \min_{s \in \Sigma,\, a \neq \alpha^{\star}(s)} |g_R(s, a; q^{\star})|\right\},\\
\bar{u}_S(\mathcal{G}, q^{\star}) &:= \max_{\theta, s, a} |u_S(\theta, s, a) - u_S(\theta, s, \alpha^{\star}(s))|, \quad \text{and}\\
\bar{u}_R(\mathcal{G}, q^{\star}) &:= \max_{\theta, s, a} |u_R(\theta, s, a) - u_R(\theta, s, \alpha^{\star}(s))|.
\end{align*}
It is not hard to observe that $\delta(\mathcal{G}, q^{\star})$ and $\bar{u}_R(\mathcal{G}, q^{\star})$ are strictly positive for a strict $\PBE$ $q^{\star}$.

\begin{restatable}{lemma}{positivityconstants}\label{lem:positivity-constants}
For a strict $\PBE$ $q^{\star} \in X^{\circ}(\mathcal{G})$, it holds that $\delta(\mathcal{G}, q^{\star}) > 0$ and $\bar{u}_R(\mathcal{G}, q^{\star}) >0$.
\end{restatable}
\begin{proof}
Since $q^{\star}$ is strict, every type $\theta \in \Theta$ has a unique best response $\sigma^{\star}(\theta)$ against $q^{\star}_R$, and therefore $g_S(\theta, s;\, q^{\star}) < 0$ holds for every $s \neq \sigma^{\star}(\theta)$. Similarly, every signal $s \in \Sigma$ has a unique best response $\alpha^{\star}(s)$ under $\mu^{\star}(\cdot \mid s)$, and therefore $g_R(s, a;\, q^{\star}) < 0$ holds for every $a \neq \alpha^{\star}(s)$. Taking the minimum of absolute values over the finite index sets gives $\delta(\mathcal{G}, q^{\star}) > 0$.

To show that $\bar{u}_R(\mathcal{G}, q^{\star}) > 0$ suppose (toward contradiction) the contrary. If $\bar{u}_R(\mathcal{G}, q^{\star}) = 0$, then $u_R(\theta, s, a) = u_R(\theta, s, \alpha^{\star}(s))$ for every $\theta, s, a$, so the receiver is indifferent between every $a \in A \setminus \{\alpha^{\star}(s)\}$ and $\alpha^{\star}(s)$ at every signal $s$. This contradicts that the strict $\PBE$ $q^{\star}$ has a unique best response for every $s \in \Sigma$.
\end{proof}

With Lemma~\ref{lem:positivity-constants} in hand, we obtain a quantitative characterization of the variational stability neighborhood of $q^{\star}$, with an explicit radius determined by $\delta(\mathcal{G}, q^{\star})$, $\bar{u}_S(\mathcal{G}, q^{\star})$, $\bar{u}_R(\mathcal{G}, q^{\star})$, and $\min_{\theta \in \Theta} \pi(\theta)$.

\begin{theorem}\label{thm:StictPBE_implies_VS}
     Let $q^{\star} \in X^{\circ}(\mathcal{G})$ be a strict $\PBE$ of $\mathcal{G}$. Let
     $
     L \;:=\; \max\left\{\bar{u}_S(\mathcal{G}, q^{\star}),\, \frac{2\bar{u}_R(\mathcal{G}, q^{\star})}{\min_{\theta \in \Theta} \pi(\theta)}\right\} \;>\; 0.
     $
     Then, for all $q \in X^{\circ}(\mathcal{G})$ with $\|q-q^{\star}\|_1 < \delta(\mathcal{G}, q^{\star}) / (2L)$, $\langle g(q),\, q-q^{\star}\rangle \leq 0$ holds, with strict inequality whenever $q \neq q^{\star}$.%
\end{theorem}

The quantity $\delta(\mathcal{G}, q^{\star}) / (2L)$ in Theorem \ref{thm:StictPBE_implies_VS} has a transparent interpretation: $\delta$ measures how strict the $\PBE$ $q^{\star}$ is, while $L$ captures how rapidly $g$ varies near $q^{\star}$: it is their ratio that sets the radius of the variational stability neighborhood. The proof of Theorem~\ref{thm:StictPBE_implies_VS} relies on two technical lemmas.%

\begin{restatable}{lemma}{strictpbestampacchia}\label{lem:strict-pbe-stampacchia}
Let $q^{\star} \in X^{\circ}(\mathcal{G})$ be a strict $\PBE$ of $\mathcal{G}$. Then for every $q \in X^{\circ}(\mathcal{G})$,
\[
\langle g(q^{\star}),\, q-q^{\star}\rangle \;\leq\; -\frac{\delta(\mathcal{G}, q^{\star})}{2}\,\|q - q^{\star}\|_1.
\]
\end{restatable}
    \begin{proof}
    Fix $q \in X^{\circ}(\mathcal{G})$. We have:
    \begin{align*}
        \langle g(q^{\star}), q-q^{\star}\rangle
        &= \sum_{\theta \in \Theta} \sum_{s \in \Sigma} g_S(\theta, s;\,q^{\star}) \bigl[q_S(s \mid \theta) - q^{\star}_S(s \mid \theta)\bigr] \\
        &\quad + \sum_{s \in \Sigma} \sum_{a \in A} g_R(s, a;\,q^{\star}) \bigl[q_R(a \mid s) - q^{\star}_R(a \mid s)\bigr] \\
        &= \sum_{\theta \in \Theta} \sum_{s \neq \sigma^{\star}(\theta)} g_S(\theta, s;\,q^{\star})\, q_S(s \mid \theta) + \sum_{s \in \Sigma} \sum_{a \neq \alpha^{\star}(s)} g_R(s, a;\,q^{\star})\, q_R(a \mid s) \\
        &\leq -\delta(\mathcal{G}, q^{\star}) \cdot \Bigl(\sum_{\theta \in \Theta} \sum_{s \neq \sigma^{\star}(\theta)} q_S(s \mid \theta) + \sum_{s \in \Sigma} \sum_{a \neq \alpha^{\star}(s)} q_R(a \mid s)\Bigr) \\
        &= -\frac{\delta(\mathcal{G}, q^{\star})}{2} \cdot \|q-q^{\star}\|_1.
    \end{align*}

        Here, the second equality follows since $q^{\star}_S(s \mid \theta) = 0$ for $s \neq \sigma^{\star}(\theta)$ and $q^{\star}_R(a \mid s) = 0$ for $a \neq \alpha^{\star}(s)$ (by definition of $\sigma^{\star}, \alpha^{\star}$), together with $g_S(\theta, \sigma^{\star}(\theta); q^{\star}) = 0$ and $g_R(s, \alpha^{\star}(s); q^{\star}) = 0$ (by our choice of reference). The inequality holds since $g_S(\theta, s; q^{\star}) \le -\delta(\mathcal{G}, q^{\star})$ for $s \neq \sigma^{\star}(\theta)$ and $g_R(s, a; q^{\star}) \le -\delta(\mathcal{G}, q^{\star})$ for $a \neq \alpha^{\star}(s)$. To see this, observe that for every $\theta \in \Theta$ and $s \neq \sigma^{\star}(\theta)$, and for every $s \in \Sigma$ and $a \neq \alpha^{\star}(s)$, respectively,
\begin{align*}
g_S(\theta, s; q^{\star}) &= u_S\bigl(\theta, s, \alpha^{\star}(s)\bigr) - u_S\bigl(\theta, \sigma^{\star}(\theta), \alpha^{\star}(\sigma^{\star}(\theta))\bigr) < 0,\\
g_R(s, a; q^{\star}) &= \sum_{\theta \in \Theta} \mu^{\star}(\theta \mid s)\, u_R(\theta, s, a) - \sum_{\theta \in \Theta} \mu^{\star}(\theta \mid s)\, u_R\bigl(\theta, s, \alpha^{\star}(s)\bigr) < 0.
\end{align*}
The equalities follow from \eqref{eq:g-sender} and \eqref{eq:g-receiver}, since $q^{\star}_R$ is pure and $(\bar{\sigma}, \bar{\alpha}) = (\sigma^{\star}, \alpha^{\star})$, and the strict inequalities are the Sender's and the Receiver's strictness conditions in the definition of a strict $\PBE$ (Section~\ref{sec:model}). As both quantities are strictly negative and $\delta(\mathcal{G}, q^{\star})$ is the minimum of their absolute values over all such pairs, the two inequalities follow (this is also the argument in the proof of Lemma~\ref{lem:positivity-constants}). The last equality follows from Claim~\ref{claim:l1-decomposition}, which we state and prove below.

\begin{claim}\label{claim:l1-decomposition}
For every $q \in X^{\circ}(\mathcal{G})$,
\[
\sum_{\theta \in \Theta} \sum_{s \neq \sigma^{\star}(\theta)} q_S(s \mid \theta) \;+\; \sum_{s \in \Sigma} \sum_{a \neq \alpha^{\star}(s)} q_R(a \mid s) \;=\; \tfrac{1}{2}\,\|q - q^{\star}\|_1.
\]
\end{claim}
\begin{proof}
Fix $\theta \in \Theta$. Since $q^{\star}$ is a strict $\PBE$, $q^{\star}_S(\cdot \mid \theta)$ is the point mass on $\sigma^{\star}(\theta)$, i.e., $q^{\star}_S(\sigma^{\star}(\theta) \mid \theta)=1$ and $q^{\star}_S(s \mid \theta)=0$ for every $s \neq \sigma^{\star}(\theta)$. Hence,
\begin{align*}
\|q_S(\cdot \mid \theta) - q^{\star}_S(\cdot \mid \theta)\|_1
&= |q_S(\sigma^{\star}(\theta) \mid \theta) - 1| + \sum_{s \neq \sigma^{\star}(\theta)} |q_S(s \mid \theta) - 0| \\
&= \bigl(1 - q_S(\sigma^{\star}(\theta) \mid \theta)\bigr) + \sum_{s \neq \sigma^{\star}(\theta)} q_S(s \mid \theta) \\
&= 2 \sum_{s \neq \sigma^{\star}(\theta)} q_S(s \mid \theta),
\end{align*}
where the first equality expands the $\ell_1$ norm componentwise, using the values of $q^{\star}_S(\cdot \mid \theta)$ above, the second drops the absolute values, since $q_S(\cdot \mid \theta) \in \Delta(\Sigma)$ gives $q_S(\sigma^{\star}(\theta) \mid \theta) \le 1$ and $q_S(s \mid \theta) \ge 0$ for every $s \in \Sigma$, and the last uses $1 - q_S(\sigma^{\star}(\theta) \mid \theta) = \sum_{s \neq \sigma^{\star}(\theta)} q_S(s \mid \theta)$, again since $q_S(\cdot \mid \theta)$ is a probability distribution. Summing over $\theta \in \Theta$,
\[
\sum_{\theta \in \Theta} \sum_{s \neq \sigma^{\star}(\theta)} q_S(s \mid \theta) \;=\; \tfrac{1}{2}\,\|q_S - q^{\star}_S\|_1,
\]
and an analogous identity for the receiver gives $\sum_{s \in \Sigma} \sum_{a \neq \alpha^{\star}(s)} q_R(a \mid s) = \tfrac{1}{2}\,\|q_R - q^{\star}_R\|_1$. Adding the two and using the decomposition $\|q - q^{\star}\|_1 = \|q_S - q^{\star}_S\|_1 + \|q_R - q^{\star}_R\|_1$ gives the claim.
\end{proof}
The proof of the Lemma follows.
\end{proof}

\begin{restatable}{lemma}{glipschitz}\label{lemma:g-lipschitz}
    Let $q^{\star} \in X^{\circ}(\mathcal{G})$ be a strict $\PBE$ of $\mathcal{G}$.
    Then, for every $q \in X^{\circ}(\mathcal{G})$,
    $
    \langle g(q)-g(q^{\star}),\, q-q^{\star} \rangle \;\leq\; L \cdot \|q-q^{\star}\|_1^2,
    $
    where $L=\max\Bigl\{\bar{u}_S(\mathcal{G}, q^{\star}),\, \frac{2\,\bar{u}_R(\mathcal{G}, q^{\star})}{\min_{\theta \in \Theta} \pi(\theta)}\Bigr\}$.
\end{restatable}
\begin{proof}
    With reference $(\bar{\sigma}, \bar{\alpha}) = (\sigma^{\star}, \alpha^{\star})$, $g_S(\theta, \sigma^{\star}(\theta); q) = 0$ and $g_R(s, \alpha^{\star}(s); q) = 0$ for every $q \in X^{\circ}(\mathcal{G})$, so $g(q) - g(q^{\star})$ vanishes at the equilibrium components and we restrict the analysis to $s \neq \sigma^{\star}(\theta)$ for the sender and $a \neq \alpha^{\star}(s)$ for the receiver.

    Fix $q \in X^{\circ}(\mathcal{G})$, $\theta \in \Theta$, and $s \in \Sigma$ with $s \neq \sigma^{\star}(\theta)$. We have:
    \begin{align*}
    &\bigl|g_S(\theta, s;\,q) - g_S(\theta, s;\,q^{\star})\bigr| \\
    &\quad= \biggl|\sum_{a \in A} u_S(\theta, s, a)\,\bigl[q_R(a\mid s) - q_R^{\star}(a\mid s)\bigr] \\
    &\qquad - \sum_{a \in A} u_S\bigl(\theta, \sigma^{\star}(\theta), a\bigr)\,\bigl[q_R(a\mid \sigma^{\star}(\theta)) - q_R^{\star}(a\mid \sigma^{\star}(\theta))\bigr]\biggr| \\
    &\quad= \biggl|\sum_{a \in A}\bigl[u_S(\theta, s, a) - u_S(\theta, s, \alpha^{\star}(s))\bigr]\,\bigl[q_R(a\mid s) - q_R^{\star}(a\mid s)\bigr] \\
    &\qquad - \sum_{a \in A}\bigl[u_S\bigl(\theta, \sigma^{\star}(\theta), a\bigr) - u_S\bigl(\theta, \sigma^{\star}(\theta), \alpha^{\star}(\sigma^{\star}(\theta))\bigr)\bigr]\,\bigl[q_R(a\mid \sigma^{\star}(\theta)) - q_R^{\star}(a\mid \sigma^{\star}(\theta))\bigr]\biggr| \\
    &\quad\leq \bar{u}_S(\mathcal{G}, q^{\star})\cdot\Bigl(\|q_R(\cdot\mid s) - q_R^{\star}(\cdot\mid s)\|_1 + \|q_R(\cdot\mid \sigma^{\star}(\theta)) - q_R^{\star}(\cdot\mid \sigma^{\star}(\theta))\|_1\Bigr) \\
    &\quad\leq \bar{u}_S(\mathcal{G}, q^{\star})\cdot\|q_R - q_R^{\star}\|_1 \;\leq\; \bar{u}_S(\mathcal{G}, q^{\star})\cdot\|q - q^{\star}\|_1. \numberthis\label{eq:sender-g-lipschtz}
\end{align*}
    The second equality holds since $\sum_{a \in A} [q_R(a \mid s) - q^{\star}_R(a \mid s)] = 0$. Then, the first inequality applies the triangle inequality and the bound $\max_a |u_S(\theta, s', a) - u_S(\theta, s', \alpha^{\star}(s'))| \le \bar{u}_S(\mathcal{G}, q^{\star})$ from the definition of $\bar{u}_S(\mathcal{G}, q^{\star})$, valid for every signal $s'$.

    To establish a similar argument for the receiver, we first prove the following Claim.
    \begin{claim}\label{claim:receiver-technical-claim}
        For every $q \in X^{\circ}(\mathcal{G})$ and every $s \in \Sigma$,
        \[
        \sum_{\theta\in\Theta} \bigl|\mu(\theta\mid s;\,q) - \mu(\theta\mid s;\,q^{\star})\bigr| \;\leq\; \frac{2\,\|q_S - q_S^{\star}\|_1}{\min_{\theta \in \Theta} \pi(\theta)}.
        \]
    \end{claim}
    \begin{proof}
        Fix $q \in X^{\circ}(\mathcal{G})$ and $s \in \Sigma$. Write $f(\theta) := \pi(\theta)\,q_S(s \mid \theta)$ and $f^{\star}(\theta) := \pi(\theta)\,q^{\star}_S(s \mid \theta)$, with $Z := \sum_\theta f(\theta) > 0$ and $Z^{\star} := \sum_\theta f^{\star}(\theta) > 0$ (both positive since $q, q^{\star} \in X^{\circ}(\mathcal{G})$). Then $\mu(\theta \mid s;\, q) = f(\theta)/Z$ and $\mu(\theta \mid s;\, q^{\star}) = f^{\star}(\theta)/Z^{\star}$.

        Since $q^{\star}$ is a strict $\PBE$, $q^{\star}_S(\cdot \mid \theta)$ has unit support at $\sigma^{\star}(\theta)$. The signal $s$ being on-path under $q^{\star}$ means there exists $\theta^{\star}_s \in \Theta$ with $\sigma^{\star}(\theta^{\star}_s) = s$, so that
        \begin{equation}\label{eq:Z-star-lower-bound}
        Z^{\star} \;=\; \sum_{\theta \in \Theta} \pi(\theta)\,q^{\star}_S(s \mid \theta) \;\geq\; \pi(\theta^{\star}_s) \;\geq\; \min_{\theta \in \Theta} \pi(\theta).
        \end{equation}

        Adding and subtracting $f(\theta)\,Z$ in the numerator,
        \begin{align*}
        \mu(\theta \mid s;\, q) - \mu(\theta \mid s;\, q^{\star})
        &\;=\; \frac{f(\theta)\,Z^{\star} - f^{\star}(\theta)\,Z}{Z\,Z^{\star}}
        \;=\; \frac{f(\theta)\,(Z^{\star} - Z) + Z\,(f(\theta) - f^{\star}(\theta))}{Z\,Z^{\star}} \\
        &\;=\; \mu(\theta \mid s;\, q)\,\frac{Z^{\star} - Z}{Z^{\star}} \;+\; \frac{f(\theta) - f^{\star}(\theta)}{Z^{\star}}.
        \end{align*}
        Applying the triangle inequality term by term and summing over $\theta$, using $\sum_{\theta \in \Theta} \mu(\theta \mid s;\, q) = 1$, and combining with $|Z^{\star} - Z| \leq \|f - f^{\star}\|_1$, $\|f - f^{\star}\|_1 \leq \|q_S - q^{\star}_S\|_1$, and~\eqref{eq:Z-star-lower-bound},
        \begin{align*}
        \sum_{\theta \in \Theta} \bigl|\mu(\theta \mid s;\, q) - \mu(\theta \mid s;\, q^{\star})\bigr|
        &\;\leq\; \frac{|Z^{\star} - Z|}{Z^{\star}} \;+\; \frac{\|f - f^{\star}\|_1}{Z^{\star}}
        \;\leq\; \frac{2\,\|f - f^{\star}\|_1}{Z^{\star}} \\
        &\;\leq\; \frac{2\,\|q_S - q^{\star}_S\|_1}{Z^{\star}}
        \;\leq\; \frac{2\,\|q_S - q^{\star}_S\|_1}{\min_{\theta \in \Theta} \pi(\theta)},
        \end{align*}
        where $|Z^{\star} - Z| \leq \|f - f^{\star}\|_1$ is the triangle inequality applied to $|\sum_{\theta \in \Theta} (f^{\star}(\theta) - f(\theta))|$, and $\|f - f^{\star}\|_1 \leq \|q_S - q^{\star}_S\|_1$ uses $\pi(\theta) \leq 1$ together with enlarging the inner sum from the single signal $s$ to all of $\Sigma$.
    \end{proof}

    Fix $q \in X^{\circ}(\mathcal{G})$, $s \in \Sigma$, and $a \in A$ with $a \neq \alpha^{\star}(s)$. Then:
    \begin{align*}
    \bigl|g_R(s, a;\,q) - g_R(s, a;\,q^{\star})\bigr|
    &= \biggl|\sum_{\theta\in\Theta} \bigl[u_R(\theta, s, a) - u_R(\theta, s, \alpha^{\star}(s))\bigr]\,\bigl[\mu(\theta\mid s;\,q) - \mu(\theta\mid s;\,q^{\star})\bigr]\biggr| \\
    &\leq \max_{\theta \in \Theta} \bigl|u_R(\theta, s, a) - u_R(\theta, s, \alpha^{\star}(s))\bigr|\cdot \sum_{\theta \in \Theta} \bigl|\mu(\theta\mid s;\,q) - \mu(\theta\mid s;\,q^{\star})\bigr| \\
    &\leq \bar{u}_R(\mathcal{G}, q^{\star}) \cdot \frac{2\,\|q_S - q_S^{\star}\|_1}{\min_{\theta \in \Theta} \pi(\theta)} \;\leq\; \frac{2\,\bar{u}_R(\mathcal{G}, q^{\star})}{\min_{\theta \in \Theta} \pi(\theta)} \cdot \|q - q^{\star}\|_1. \numberthis\label{eq:receiver-g-lipschtz}
    \end{align*}
    The first inequality is the triangle inequality; the second uses the definition of $\bar{u}_R(\mathcal{G}, q^{\star})$ and Claim~\ref{claim:receiver-technical-claim}.

    Set $L := \max\bigl\{\bar{u}_S(\mathcal{G}, q^{\star}),\, 2\bar{u}_R(\mathcal{G}, q^{\star})/\min_{\theta \in \Theta} \pi(\theta)\bigr\}$. From~\eqref{eq:sender-g-lipschtz} and~\eqref{eq:receiver-g-lipschtz},
    \begin{align*}
    \|g(q) - g(q^{\star})\|_\infty &=\; \max_{\theta \in \Theta,\, s \in \Sigma,\, a \in A}\Bigl\{\bigl|g_S(\theta, s;\,q) - g_S(\theta, s;\,q^{\star})\bigr|,\; \bigl|g_R(s, a;\,q) - g_R(s, a;\,q^{\star})\bigr|\Bigr\}
    \\ &\leq\; L \cdot \|q-q^{\star}\|_1.
    \end{align*}
    We now use H\"older's inequality to obtain
    \[
    \langle g(q)-g(q^{\star}),\, q-q^{\star} \rangle \;\leq\; \|g(q) - g(q^{\star})\|_\infty \cdot \|q-q^{\star}\|_1 \;\leq\; L \cdot \|q-q^{\star}\|_1^2
    \]
    and complete the proof.
\end{proof}

We are now ready to prove Theorem~\ref{thm:StictPBE_implies_VS}.

\begin{proof}[Proof of Theorem~\ref{thm:StictPBE_implies_VS}]
    By Lemma~\ref{lem:positivity-constants}, we have that $\delta(\mathcal{G}, q^{\star}) > 0$ and $\bar{u}_R(\mathcal{G}, q^{\star}) > 0$, so $L > 0$ as claimed and the radius $\delta(\mathcal{G}, q^{\star})/(2L)$ is well-defined and positive. The case $q = q^{\star}$ is immediate and we focus on the $q \neq q^{\star}$ case to complete the proof.

    Fix $q \in X^{\circ}(\mathcal{G})$ with $0 < \|q-q^{\star}\|_1 < \delta(\mathcal{G}, q^{\star})/(2L)$. Then
    \begin{align*}
       \langle g(q),\, q-q^{\star}\rangle
       &\;=\; \langle g(q^{\star}),\, q-q^{\star}\rangle \;+\; \langle g(q) - g(q^{\star}),\, q-q^{\star}\rangle \\
       &\;\leq\; -\frac{\delta(\mathcal{G}, q^{\star})}{2}\,\|q-q^{\star}\|_1 \;+\; L\,\|q-q^{\star}\|_1^2 \\
       &\;=\; \|q-q^{\star}\|_1 \left(-\frac{\delta(\mathcal{G}, q^{\star})}{2} \;+\; L\,\|q-q^{\star}\|_1\right) \\
       &\;<\; \|q-q^{\star}\|_1 \left(-\frac{\delta(\mathcal{G}, q^{\star})}{2} \;+\; L \cdot \frac{\delta(\mathcal{G}, q^{\star})}{2L}\right) \;=\; 0,
    \end{align*}
    where the first inequality follows by Lemma~\ref{lem:strict-pbe-stampacchia}  and Lemma~\ref{lemma:g-lipschitz}. Finally, the strict inequality holds since, by assumption, $\|q-q^{\star}\|_1 < \delta(\mathcal{G}, q^{\star})/(2L)$.
\end{proof}

We conclude this section by showing that variational stability of a point $q^{\star} \in X^{\circ}(\mathcal{G})$ implies that $q^{\star}$ is a strict $\PBE$. This establishes our characterization result and completes the proof of Theorem \ref{thm:StrictPBE_equiv_VariationStable}.

\begin{restatable}{theorem}{vsimpliesstrictpbe}\label{thm:VS_implies_StrictPBE}
    Let $q^{\star} \in X^{\circ}(\mathcal{G})$. If $q^{\star}$ is variationally stable, then it is a strict $\PBE$.%
\end{restatable}
\begin{proof}
Since $q^{\star}$ is variationally stable, by Definition \ref{def:VS} there exists an open neighborhood $H \subseteq X^{\circ}(\mathcal{G})$ such that $\langle g(q), q-q^{\star} \rangle \leq 0$ for every $q \in H$, with strict inequality whenever $q \neq q^{\star}$. Because $H$ is open in $X^{\circ}(\mathcal{G})$ (which is itself open in $X(\mathcal{G})$ by Lemma \ref{lem:Xcirc-convex-open}), there exists an $\varepsilon > 0$ such that $\{q \in X(\mathcal{G}): \|q-q^{\star}\|_1 < \varepsilon \}  \subseteq H$.

We first consider the Sender. Fix $\theta_k \in \Theta$ and
$s \in \supp(q^{\star}_S(\cdot \mid \theta_k))$. For any
$s' \in \Sigma$ with $s' \neq s$, let
\[
\lambda_{\max} := \min\!\left\{q^{\star}_S(s \mid \theta_k),\,
1 - q^{\star}_S(s' \mid \theta_k),\,
\frac{\sum_{\theta \in \Theta} \pi(\theta)\, q^{\star}_S(s \mid \theta)}{\pi(\theta_k)},\,
\frac{\varepsilon}{2}\right\}.
\]
For $\lambda \in (0, \lambda_{\max})$, define $q_\lambda$ by
\[
q_{\lambda, S}(s \mid \theta_k) := q^{\star}_S(s \mid \theta_k) - \lambda,
\qquad q_{\lambda, S}(s' \mid \theta_k) := q^{\star}_S(s' \mid \theta_k) + \lambda,
\]
with all other entries of $q_{\lambda, S}$ and all of $q_{\lambda, R}$
equal to those of $q^{\star}$. The four terms in $\lambda_{\max}$ ensure, respectively, that $q_{\lambda, S}(s \mid \theta_k) \geq 0$, that $q_{\lambda, S}(s' \mid \theta_k) \leq 1$, that $q_\lambda \in X^{\circ}(\mathcal{G})$ (see (i) below), and that $\|q_\lambda - q^{\star}\|_1 = 2\lambda < \varepsilon$, so that $q_\lambda \in H$ (see (ii) below). Then:
\begin{enumerate}[label=(\roman*)]
    \item $q_\lambda \in X^{\circ}(\mathcal{G})$: by the definition of
    $\lambda_{\max}$,
    $$\sum_{\theta \in \Theta} \pi(\theta)\, q_{\lambda, S}(s \mid \theta)
    = \sum_{\theta \in \Theta} \pi(\theta)\, q^{\star}_S(s \mid \theta) - \lambda \pi(\theta_k) > 0,$$
    while $\sum_{\theta} \pi(\theta)\, q_{\lambda, S}(s' \mid \theta)
    > \sum_{\theta} \pi(\theta)\, q^{\star}_S(s' \mid \theta) > 0$,
    since $q^{\star} \in X^{\circ}(\mathcal{G})$.
    \item $q_\lambda \neq q^{\star}$ and
    $\|q_\lambda - q^{\star}\|_1 = 2\lambda < \varepsilon$.
    \item By construction of $q_\lambda$ and the fact that $g_S$ depends
    on $q$ only through $q_R$, with $q_{\lambda, R} = q^{\star}_R$,
    \[
    \langle g(q_\lambda),\, q_\lambda - q^{\star} \rangle
    = \lambda\bigl(g_S(\theta_k, s';\, q^{\star}) - g_S(\theta_k, s;\, q^{\star})\bigr).
    \]
\end{enumerate}
By (i)--(ii) and strict variational stability of $q^{\star}$,
$\langle g(q_\lambda),\, q_\lambda - q^{\star} \rangle < 0$. Combined
with (iii), this gives
\begin{equation}
g_S(\theta_k, s';\, q^{\star}) < g_S(\theta_k, s;\, q^{\star})
\qquad \text{for every } s' \in \Sigma \setminus \{s\}.
\label{eq:sender-strict}
\end{equation}
Since \eqref{eq:sender-strict} holds for every
$s \in \supp(q^{\star}_S(\cdot \mid \theta_k))$, the support is a
singleton: if it contained two distinct signals $s, s'$, applying
\eqref{eq:sender-strict} to $(s, s')$ and to $(s', s)$ would yield
$g_S(\theta_k, s';\, q^{\star}) < g_S(\theta_k, s;\, q^{\star})
< g_S(\theta_k, s';\, q^{\star})$, a contradiction. Writing
$\sigma^{\star}(\theta_k)$ for the unique element,
\eqref{eq:sender-strict} says $\sigma^{\star}(\theta_k)$ is the
strict best response for type $\theta_k$ against $q^{\star}_R$.

By the analogous argument with
$a \in \supp(q^{\star}_R(\cdot \mid s))$, $a' \in A \setminus \{a\}$,
and $\lambda_{\max} := \min\{q^{\star}_R(a \mid s),\, 1 - q^{\star}_R(a' \mid s),\, \varepsilon/2\}$
(here membership in $X^{\circ}(\mathcal{G})$ is automatic since
$q_{\lambda, S} = q^{\star}_S$): for every $s \in \Sigma$,
$\supp(q^{\star}_R(\cdot \mid s))$ is a singleton $\{\alpha^{\star}(s)\}$,
and $\alpha^{\star}(s)$ is the strict best response under the Bayes
posterior $\mu^{\star}(\cdot \mid s)$.

Bayes consistency holds since $q^{\star} \in X^{\circ}(\mathcal{G})$.
Therefore $q^{\star}$ is a strict $\PBE$.
\end{proof}

\section{Global Average-iterate Convergence}\label{sec:convergence}

Recall that in Theorem~\ref{thm:negative-K2} (Section~\ref{subsec:limitations}) we present a simple finite signaling game with a strict $\PBE$ at which the Minty condition fails (and hence monotonicity also fails; see Definition~\ref{def:variational-conditions}). A feature of this instance is that the sender has a dominant strategy to  signal her type, while the receiver has a unique best response upon receiving an ``honest'' signal. In this section we abstract these features into two assumptions on the signaling game and show that, under them, the unique $\PBE$ (the Riley assessment) is in fact the unique $\EFBCCE$. As we discuss below (Corollary~\ref{cor:avg-iterate-Riley}), this yields a \emph{global} average-iterate convergence guarantee to the unique $\PBE$ for this class of signaling games, which is determined by Assumptions \ref{assum:receiver-unique-best-response} and \ref{assum:sender-dominance} below. We call signaling games satisfying both assumptions \emph{honest-signaling games}.

\begin{assumption}\label{assum:receiver-unique-best-response}
    For every $k \in \{1, \ldots, K\}$ and every $j\neq k$, $u_R(\theta_k, s_k, a_k) > u_R(\theta_k, s_k, a_j)$.
\end{assumption}

\begin{assumption}\label{assum:sender-dominance}
    For every $k \in \{1,\dots, K\}$ and every $j \neq k$,
     $\min_{a \in A}\,u_S(\theta_k, s_k, a) > \max_{a \in A}\,u_S(\theta_k, s_j, a).$

\end{assumption}
Assumption~\ref{assum:receiver-unique-best-response} says that when the receiver believes the sender's type is $\theta_k$, action $a_k$ is the unique best response at the truthful signal $s_k$. This unique-best-response condition is standard in the signaling-games literature\footnote{In continuous-action games, an analogous role is played by the assumption that the receiver's pointwise utility is strictly quasi-concave in the action, see e.g., \citep{sobel2020signaling}.} \citep{cho1987, sobel2020signaling, mailath1987incentive}. Assumption~\ref{assum:sender-dominance} says that type $\theta_k$ strictly prefers the truthful signal $s_k$ over any non-truthful signal $s_j$, regardless of the receiver's action. The main result of this section is Theorem \ref{thm:unique-EFCCE}.%

\begin{restatable}{theorem}{uniqueBEFCE}\label{thm:unique-EFCCE}
    For every honest-signaling game,
    the Riley assessment $(q^{\star}, \mu^{\star})$ is the unique $\PBE$. Moreover, $q^{\star}$ is the unique $\EFBCCE$.
\end{restatable}

In Lemma \ref{lem:riley-pbe} we show that the Riley outcome is supported on a $\PBE$.

\begin{lemma}\label{lem:riley-pbe}
    Under Assumptions~\ref{assum:receiver-unique-best-response}
    and~\ref{assum:sender-dominance},
    the Riley outcome $q^{\star}$ is a $\PBE$ strategy.
\end{lemma}
\begin{proof}
We construct the belief system $\mu^{\star}$ associated with $q^{\star}$ via Bayes' rule and show that the assessment $(q^{\star}, \mu^{\star})$ satisfies the conditions of Definition~\ref{def. Perfect Bayesian Equilibrium} (consistency of beliefs and sequential rationality for both players), so that $q^{\star} \in \SPBE$.

    Every signal $s_k \in \Sigma$ satisfies $\sum_{\theta \in \Theta}\pi(\theta)q_S^{\star}(s_k \mid \theta)=\pi(\theta_k)>0$. Hence $q^{\star} \in X^{\circ}(\mathcal{G})$ and, by \eqref{eq. posterior update}, it must be that
    \begin{equation}\label{eq:bayes-sep}
    \mu^\star(\theta \mid s_k)
    = \frac{\pi(\theta)\,q_S^\star(s_k\mid\theta)}
           {\sum_{\theta'}\pi(\theta')\,q_S^\star(s_k\mid\theta')}
    = \frac{\pi(\theta)\,\mathbf{1}\{\theta=\theta_k\}}{\pi(\theta_k)}
    = \mathbf{1}\{\theta=\theta_k\},
    \end{equation}
which holds by construction. Hence, consistency of beliefs follows.

We now show sequential rationality per player. For the receiver, we have that $\supp(q_R^{\star}(\cdot \mid s_k)) = \{a_k\}$ holds for every $k \in \{1, \ldots, K\}$.  We need to show that

\begin{equation}\label{eq:seq-sender}
   a_k \in \arg\max_{a \in A} \sum_{\theta \in \Theta} \mu^{\star}( \theta \mid s_k) u_R(\theta, s_k, a)\, q_R^\star(a \mid s).
\end{equation}%

Fix $k \in \{1,\dots, K\}$. For every $j \in \{1, \ldots, K\}\setminus \{k\}$,
\begin{align*}
&\sum_{i=1}^K \mu^\star(\theta_i \mid s_k)\, u_R(\theta_i, s_k, a_k)
= \sum_{i=1}^K \mathbf{1}[\theta_i = \theta_k]\, u_R(\theta_i, s_k, a_k) = u_R(\theta_k, s_k, a_k) \\
&> u_R(\theta_k, s_k, a_j) = \sum_{i=1}^K \mathbf{1}[\theta_i = \theta_k]\, u_R(\theta_i, s_k, a_j)
= \sum_{i=1}^K \mu^\star(\theta_i \mid s_k)\, u_R(\theta_i, s_k, a_j).
\end{align*}
The first and last equalities are by~\eqref{eq:bayes-sep} and the strict inequality is Assumption~\ref{assum:receiver-unique-best-response}. Hence \eqref{eq:seq-sender} holds.

For the sender, $\supp(q_S^\star(\cdot \mid \theta_k)) = \{s_k\}$ for every $k \in \{1, \ldots, K\}$. We need to show that

\begin{equation}\label{eq:seq-rational-pbe-riley}
   s_k \in \arg\max_{s \in \Sigma} \sum_{a \in A} u_S(\theta_k, s, a)\, q_R^\star(a \mid s).
\end{equation}

Fix $k \in \{1, \ldots, K\}$. For every $j \in \{1, \ldots, K\} \setminus \{k\}$, we have that
\begin{align*}
&\sum_{a \in A} u_S(\theta_k, s_k, a)\, q_R^\star(a \mid s_k)
= \sum_{a \in A} u_S(\theta_k, s_k, a)\, \mathbf{1}[a = a_k] = u_S(\theta_k, s_k, a_k) \\&\ge \min_{a \in A} u_S(\theta_k, s_k, a) > \max_{a \in A} u_S(\theta_k, s_j, a)
\ge u_S(\theta_k, s_j, a_j) \\
&= \sum_{a \in A} u_S(\theta_k, s_j, a)\, \mathbf{1}[a = a_j]
= \sum_{a \in A} u_S(\theta_k, s_j, a)\, q_R^\star(a \mid s_j).
\end{align*}
Here, the first and last equalities are by the definition of $q_R^\star$.
Finally, the strict inequality is Assumption~\ref{assum:sender-dominance}. \eqref{eq:seq-rational-pbe-riley} follows and this concludes the proof.
\end{proof}

We now show that the Riley assessment $(q^{\star}, \mu^{\star})$ is the unique $\PBE$. In fact we prove a stronger statement in Lemma~\ref{lem:unique-BEFCE}: $(q^{\star}, \mu^{\star})$ is the unique $\EFBCCE$ of $\mathcal{G}$.\footnote{Recall that by the hierarchy of solution concepts, $\SPBE \subseteq \SEFBCCE$. Therefore $q^{\star} \in \EFBCCE$ follows immediately from Lemma~\ref{lem:riley-pbe}.}

\begin{lemma}\label{lem:unique-BEFCE}
    For every signaling game $\mathcal{G}$ satisfying  Assumptions~\ref{assum:receiver-unique-best-response}
    and~\ref{assum:sender-dominance}, the Riley assessment $(q^{\star}, \mu^{\star})$ is the unique $\EFBCCE$ in $X(\mathcal{G})$.
\end{lemma}
\begin{proof}
Let $Q \in \Delta(\Sigma^{\Theta} \times A^{\Sigma})$
be any $\EFBCCE$. We show that $Q = (q_S^{\star}, q_R^{\star})$ (see \eqref{eq:riley}). Fix $k \in \{1, \ldots, K\}$. By invoking~\eqref{eq:efcce-sender} with $\theta = \theta_k$ and $s' = s_k$, we obtain
\begin{equation}\label{eq:BEFCE-sender-dev}
\mathbb{E}_{(\sigma,\alpha) \sim Q}\bigl[
  u_S(\theta_k, \sigma(\theta_k), \alpha(\sigma(\theta_k)))
\bigr]
\;\ge\;
\mathbb{E}_{(\sigma,\alpha) \sim Q}\bigl[
  u_S(\theta_k, s_k, \alpha(s_k))
\bigr].
\end{equation}
For every realization $(\sigma, \alpha)$, Assumption~\ref{assum:sender-dominance} implies $u_S(\theta_k, s_k, \alpha(s_k)) \ge u_S(\theta_k, \sigma(\theta_k), \alpha(\sigma(\theta_k)))$, with strict inequality whenever $\sigma(\theta_k) \neq s_k$. Indeed, the two sides coincide if $\sigma(\theta_k) = s_k$, while if $\sigma(\theta_k) = s_j$ for some $j \neq k$, then
\[
u_S(\theta_k, s_k, \alpha(s_k)) \;\ge\; \min_{a \in A} u_S(\theta_k, s_k, a) \;>\; \max_{a \in A} u_S(\theta_k, s_j, a) \;\ge\; u_S(\theta_k, s_j, \alpha(s_j)).
\]
Hence, if $\Pr_{(\sigma, \alpha)\sim Q}[\sigma(\theta_k) \neq s_k] > 0$, taking expectations gives the strict reverse of~\eqref{eq:BEFCE-sender-dev}, a contradiction. We conclude that $\Pr_{(\sigma, \alpha)\sim Q}[\sigma(\theta_k) = s_k] = 1 = q^{\star}_S(s_k \mid \theta_k)$. Since $k$ was arbitrary, every signal $s \in \Sigma$ is on-path. Therefore, by \eqref{eq:bayes-sep} we also have, for every $\theta \in \Theta$ and every $k \in \{1,\dots, K\}$ that:
\begin{equation*}
  \mu(\theta \mid s_k)
= \frac{\pi(\theta)\,\Pr_{(\sigma,\alpha) \sim Q}[\sigma(\theta) = s_k]}
       {\sum_{\theta' \in \Theta} \pi(\theta')\,\Pr_{(\sigma,\alpha) \sim Q}[\sigma(\theta') = s_k]}= \frac{\pi(\theta)\,\mathbf{1}\{\theta = \theta_k\}}
       {\pi(\theta_k)} = \mathbf{1}\{\theta = \theta_k\}.
\end{equation*}
Fix $k \in \{1, \ldots, K\}$. Substituting into~\eqref{eq:efcce-receiver}, for every $a' \in A$,
\begin{align*}
\mathbb{E}_{(\sigma,\alpha) \sim Q}\bigl[u_R(\theta_k, s_k, \alpha(s_k))\bigr]
&= \sum_{\theta \in \Theta} \mu(\theta \mid s_k)\, \mathbb{E}_{(\sigma,\alpha) \sim Q}\bigl[u_R(\theta, s_k, \alpha(s_k)) \mid \sigma(\theta) = s_k\bigr]\\
&\ge \sum_{\theta \in \Theta} \mu(\theta \mid s_k)\, u_R(\theta, s_k, a')
\\
&= u_R(\theta_k, s_k, a').
\end{align*}
Setting $a' = a_k$ and applying Assumption~\ref{assum:receiver-unique-best-response} (which gives $u_R(\theta_k, s_k, a_k) > u_R(\theta_k, s_k, a)$ for every $a \neq a_k$) forces $\alpha(s_k) = a_k$ almost surely. Hence $Q = (q_S^{\star}, q_R^{\star})$ and the proof follows.
\end{proof}

By combining Lemmata \ref{lem:riley-pbe} and \ref{lem:unique-BEFCE} the proof of Theorem \ref{thm:unique-EFCCE} follows since every honest-signaling game satisfies Assumptions \ref{assum:receiver-unique-best-response} and \ref{assum:sender-dominance}.

Trigger regret captures the cumulative payoff gain a player could obtain by ``triggering'' a deviation that may depend on the recommendation; coarse-trigger regret restricts the deviation to be fixed before the recommendation is seen. Algorithms that minimize trigger regret converge in average iterate to the $\EFBCE$\footnote{In a (Bayesian) extensive-form correlated equilibrium ($\EFBCE$)~\citep{vonstengel08}, the player observes the recommendation before deciding whether to follow it; the deviation may then depend on the recommendation. For this reason, it holds that $\SEFBCE \subseteq \SEFBCCE$.} polytope~\citep{farina2022simple_jacm}, while algorithms that minimize coarse-trigger regret converge to the (broader) $\EFBCCE$ polytope~\citep{AnagnostidesFarinaKroerCelliSandholm2022}. Since Theorem~\ref{thm:unique-EFCCE} shows that the $\EFBCCE$ polytope is a singleton for this class of games, we obtain the following corollary.

\begin{corollary}\label{cor:avg-iterate-Riley}
Let $\mathcal{G}$ be an honest-signaling game. If both the Sender and the Receiver run algorithms which minimize coarse-trigger regret, the empirical distribution of play converges in average iterate to the Riley assessment $(q^{\star}, \mu^{\star})$.
\end{corollary}

Concrete examples include the trigger-regret-minimizing dynamics of \citet{farina2022simple_jacm} and the coarse-trigger-regret-minimizing algorithms of \citet{AnagnostidesFarinaKroerCelliSandholm2022, FarinaCelliGattiSandholm2021}.

\section{Experiments}\label{sec:experiments}
Our experiments study whether simple first-order learning dynamics converge to
designated equilibrium targets in small signaling games. The experiments are implemented in the
behavioral strategy space of binary signaling games and are designed to
complement the theoretical analysis. We focus on $K=2$ and extend two benchmarks to $K=4$ in Section \ref{sec:Beyond K=2}. This case is already rich enough to yield multiple equilibrium classes and multiple equilibria. We focus on the equilibrium selection problem, which is predominantly concerned with resolving multiplicity rather than the problem's dimensionality. This setting provides a suitable baseline for the analysis.
\subsection{Experimental setup}
We study learning dynamics in benchmark $2{\times}2{\times}2$ signaling games classified \emph{endogenously} by their equilibrium structure. The seven families span the qualitatively distinct cases
that arise in two-type signaling games.
Three families admit a unique pure-strategy $\PBE$: a \emph{unique separating} family in which each type sends a distinct signal that perfectly reveals
private information; a \emph{unique pooling at the low signal} and \emph{unique pooling at the high signal}. Two further families
admit multiple pure $\PBE$: a Spence-style \emph{pooling-and-separating}
family modeled on the job-market signaling problem of
\citet{spence1978job}, in which a pooling equilibrium and a
separating equilibrium coexist without belief refinements
\citep[e.g.\ the Intuitive Criterion of][]{cho1987}; and a
\emph{multiple separating}, in which two strict pure separating
profiles -- the Riley outcome and its permutation
coexist as $\PBE$. A \emph{multiple-pooling} family admits several pure pooling profiles
but no separating equilibrium. Finally, a \emph{unique
semi-separating} (see Definition \ref{def. signaling games PBE}) family modeled on the reputation game of
\citet{kreps1982reputation} has a strictly mixed unique $\PBE$ in which
one sender type plays a pure strategy and the other mixes -- the
only family in our set whose equilibrium is not pure \footnote{Other families may admit mixed-strategy equilibria as well; what singles out the semi-separating family is the absence of any pure-strategy equilibrium.}. For more details on the games, we refer to Appendix \ref{app:families}.
\Cref{tab:treatments} lists the hyperparameters per
family.
Across families we compare four learning algorithms -- projected
gradient ascent (PGA) \citep{wang2013simplex}, entropic online
mirror ascent (OMA), optimistic entropic online mirror ascent
(Opt.\,OMA) \citep{syrgkanis2015fast}, and proximal policy
optimisation (PPO) \citep{schulman2017ppo} on a tabular
policy/value parameterization -- and consider both local and global initializations. For more details on the setting and treatment variables, we refer to Appendix \ref{subsec: TreamentVariables}

The simulation design has two stages. First, we generate a candidate signaling game from a parameterized family intended to represent the corresponding equilibrium class. We retain only parameter configurations whose realized equilibrium class matches the intended benchmark family. Second, for each retained game, we simulate learning from randomized initial conditions under the selected algorithm and hyperparameters.

Convergence is evaluated relative to the equilibrium set associated with the classified game. In unique-equilibrium cases, the target is a singleton; in multiple-pure games, the target is the full set of pure equilibria, so a run counts as successful if it approaches any one of them. For each
treatment combination we report convergence frequency
($\textsc{Conv.}$), average terminal distance to the target set
($\textsc{Dist.}$), and median hitting time ($\textsc{Hit}$) over
$n=50$ paired-init seeds. Precisely, \textsc{Dist.} evaluates the Frobenius norm of the difference between the realised agent strategy and the (closest) target equilibrium strategy for the agent. For pooling targets, the receiver's response to the off-path signal is a free parameter of the equilibrium and is excluded. $\textsc{Hit}$ is reported in algorithm-internal rounds, and a PPO round packages a
  $512$-episode rollout plus $16$ mini-batch gradient updates per agent, whereas a PGA
  / OMA / Opt.OMA round is one closed-form expected-payoff gradient step.
Convergence is defined as ending within
$\delta=10^{-2}$ of the relevant equilibrium set; distance measures
residual terminal error; hitting time records the first iteration at
which the trajectory enters the tolerance neighborhood, and is
computed only over runs that converge.

\subsection{Results}

\input{cross_tab}

Table \ref{tab:endogenous_constant_results} shows that all four
algorithms perform very well in games with a unique pure
equilibrium. For unique separating and the two unique pooling
benchmarks (pooling at $m_0$ and at $m_1$), convergence is
essentially perfect under both local and global initialization,
although the mirror-based methods typically require more iterations
from global starting points. This indicates that once the
equilibrium structure is simple and the target is unique, all four
learning rules locate the relevant equilibrium set reliably, with
PGA generally reaching it fastest, OMA and Opt.\,OMA moving more
gradually and PPO being the slowest among them. This pattern also holds for the Spence-style signaling game and the multiple separating regime. The algorithms reliably reach an equilibrium, but the equilibrium depends on the basin of the initial condition rather than on the algorithm. Interestingly, our instances in the full-convergence class attain strict equilibria, which aligns with our theoretical results and demonstrates positive outcomes even beyond the formal guarantees.

Performance is weaker in the multiple-pooling and unique
semi-separating environments, but for different reasons. In the
multiple-pooling case, convergence remains high overall, yet it is
somewhat lower under local initialization for the three classical
learners than under global initialization (Opt.\,OMA: $0.48$ local
vs.\ $0.84$ global; OMA: $0.52$ vs.\ $0.82$; PGA: $0.56$ vs.\
$0.82$). A natural explanation is that ``local'' initialization is
defined relative to the first profile in the family's pure-pooling
target set, so some starting points are placed near boundary regions
associated with one equilibrium but not necessarily inside a strong
basin of attraction. In these cases, the learner can become trapped
near a nearby non-equilibrium boundary configuration or drift toward
the wrong corner region before entering the tolerance neighborhood of
any pure equilibrium. By contrast, global initialization sometimes
places the process in regions from which the overall drift more
clearly selects one of the pure equilibria, which can improve
eventual convergence even though the hitting times are much longer.
PPO is the exception in this regard. \\
In the unique semi-separating benchmark
\citep{kreps1982reputation}, the learning dynamics rarely settle
near the target equilibrium for three of the four algorithms.
Instead of converging, most trajectories under PGA, OMA, and PPO
continue to move through the state space for the full horizon: the
state does not become nearly stationary, the learner keeps updating
and exploring different regions, yet it does not enter the
tolerance neighborhood of the semi-separating equilibrium.
The sole exception is Opt.~OMA, which converges for \(100\%\) of seeds under
both local and global initialization
with median hitting times of \(40.5\) and \(70\) rounds, respectively.

These results identify the optimistic look-ahead step as a decisive algorithmic
ingredient for convergence to the mixed semi-separating equilibrium.
Understanding when and why optimistic
methods select equilibria in signaling games and whether this
extends to broader classes of extensive-form games is a promising
direction for future research.

\subsection{Scaling to Four Types}\label{sec:Beyond K=2}

We now move from $2{\times}2{\times}2$ to $4{\times}4{\times}4$ signaling games (four types, signals and actions) to test whether our observations are driven by the equilibrium structure rather than by the dimension of the problem.
Classifying games by their equilibrium structure becomes considerably harder in this setting. We therefore focus on two benchmarks with a provably unique $\PBE$, one from each convergence class of Table~\ref{tab:endogenous_constant_results}. In particular, we choose a four-type extension of the honest-signaling game, whose unique $\PBE$ is strict and separating, and a multi-threat extension of the reputation game of \citet{kreps1982reputation}, in which three types each insist on their own signal and the strategic type mixes over all four signals, giving a unique semi-separating $\PBE$. More details on these game families are given in Appendix~\ref{app:k4-benchmarks}. Algorithms, step sizes, initialization schemes, and the $n=50$ paired seeds are those of Table~\ref{tab:endogenous_constant_results}.

\begin{table}[t]
\centering
\caption{Aggregate results for the four-type benchmarks of Appendix~\ref{app:k4-benchmarks} ($K=4$), with the step sizes, initializations and $n=50$ paired seeds of Table~\ref{tab:endogenous_constant_results}. Horizon: $T=2000$ ($240$ PPO rounds) for unique separating and $T=12000$ ($1440$ PPO rounds) for unique semi-separating.}
\label{tab:k4_results}
\setlength{\tabcolsep}{4pt}
\begin{tabular}{l cccc cccc}
\toprule
 & \multicolumn{4}{c}{Local init.} & \multicolumn{4}{c}{Global init.} \\
\cmidrule(lr){2-5}\cmidrule(lr){6-9}
 & PGA & OMA & Opt.\,OMA & PPO & PGA & OMA & Opt.\,OMA & PPO \\
\midrule
\multicolumn{9}{l}{\textsc{Hit} --- full convergence class} \\
Unique separating            & 1 & 2 & 1 & 9 & 5 & 7 & 5 & 17 \\
\midrule
\multicolumn{9}{l}{Partial convergence class} \\
Semi-sep.\ \textsc{Conv.}    & .000  & .000  & 1.000  & .000  & .000  & .000  & .820 & .000  \\
Semi-sep.\ \textsc{Dist.}    & 1.335 & 1.831 & .009   & 1.904 & 1.346 & 1.820 & .201 & 1.896 \\
Semi-sep.\ \textsc{Hit}      & --    & --    & 9936.5 & --    & --    & --    & 201  & --    \\
\bottomrule
\end{tabular}
\end{table}

\Cref{tab:k4_results} shows that the qualitative picture of the binary case carries over. In the unique separating game, all four algorithms converge in every run from both initializations, with hitting times close to those for $K=2$; only PPO slows down noticeably ($9$ and $17$ rounds vs.\ $3$ and $11$). In the semi-separating game, Opt.\,OMA is again the only algorithm that converges, while PGA, OMA and PPO never enter the tolerance neighborhood and end far from the equilibrium. The main caveat is speed. Under the horizon $T=2000$ of Table~\ref{tab:endogenous_constant_results}, no Opt.\,OMA run from a local initialization has converged yet, so we extend the horizon to $T=12000$. All local runs then converge, but they first enter the tolerance neighborhood only after $6{,}033$ to $10{,}630$ iterations, compared with a median of $40.5$ for $K=2$. Under global initialization, $82\%$ of the runs converge. Half of all runs do so within $400$ iterations, the remaining converging runs take several thousand, and the other $18\%$ stay far from the equilibrium.
Thus, a larger dimension makes convergence to the mixed equilibrium much slower, but it does not change which algorithms succeed.

\section*{Acknowledgments}
This project has received funding from the European Research Council (ERC) under the European Union’s Horizon Europe research and innovation program (grant agreement No 101198689).
The authors declare no competing interests.

\bibliographystyle{plainnat}
\bibliography{signaling}
\newpage
\appendix


\section{Independence from the Reference Profile in \texorpdfstring{$g$}{g}}\label{app:reference-invariance}
\referenceinvariance*
\begin{proof}
By Definition~\ref{def:PayoffDifferenceField}, for every $\theta \in \Theta$ and $s \in \Sigma$,
\begin{align*}
g^{(1)}_S(\theta, s;\, q) - g^{(2)}_S(\theta, s;\, q)
&= \sum_{a' \in A} u_S\bigl(\theta, \bar{\sigma}_2(\theta), a'\bigr)\,q_R\bigl(a' \mid \bar{\sigma}_2(\theta)\bigr) \\
&\quad - \sum_{a' \in A} u_S\bigl(\theta, \bar{\sigma}_1(\theta), a'\bigr)\,q_R\bigl(a' \mid \bar{\sigma}_1(\theta)\bigr) \;=:\; c_S(\theta;\, q),
\end{align*}
which depends only on $\theta$ and $q$ (not on $s$). Therefore,
\begin{align*}
\langle g^{(1)}_S(q'),\, q_S - q'_S\rangle - \langle g^{(2)}_S(q'),\, q_S - q'_S\rangle
&= \sum_{\theta \in \Theta} \sum_{s \in \Sigma} c_S(\theta;\, q')\bigl[q_S(s \mid \theta) - q'_S(s \mid \theta)\bigr] \\
&= \sum_{\theta \in \Theta} c_S(\theta;\, q')\sum_{s \in \Sigma} \bigl[q_S(s \mid \theta) - q'_S(s \mid \theta)\bigr] \\
&= \sum_{\theta \in \Theta} c_S(\theta;\, q')\cdot 0 \;=\; 0,
\end{align*}
since $q_S(\cdot \mid \theta), q'_S(\cdot \mid \theta) \in \Delta(\Sigma)$ for every $\theta$.

Analogously for the receiver, for every $s \in \Sigma$ and $a \in A$,
\[
g^{(1)}_R(s, a;\, q) - g^{(2)}_R(s, a;\, q) \;=\; \sum_{\theta \in \Theta} \bigl[u_R\bigl(\theta, s, \bar{\alpha}_2(s)\bigr) - u_R\bigl(\theta, s, \bar{\alpha}_1(s)\bigr)\bigr]\,\mu(\theta \mid s) \;=:\; c_R(s;\, q),
\]
which depends only on $s$ and $q$ (not on $a$). The same argument as above gives
\[
\langle g^{(1)}_R(q'),\, q_R - q'_R\rangle - \langle g^{(2)}_R(q'),\, q_R - q'_R\rangle
\;=\; \sum_{s \in \Sigma} c_R(s;\, q')\sum_{a \in A} \bigl[q_R(a \mid s) - q'_R(a \mid s)\bigr] \;=\; 0,
\]
since $q_R(\cdot \mid s), q'_R(\cdot \mid s) \in \Delta(A)$ for every $s$.

Summing the sender and receiver identities gives $\langle g^{(1)}(q'),\, q - q'\rangle = \langle g^{(2)}(q'),\, q - q'\rangle$.
\end{proof}
Each of the three conditions in Definition~\ref{def:variational-conditions} and Definition~\ref{def:VS} is stated through inner products of the form $\langle g(q'),\, q - q'\rangle$:
\begin{itemize}
    \item Monotonicity: $\langle g(q) - g(q'),\, q - q'\rangle \leq 0$ for all $q, q' \in X^{\circ}(\mathcal{G})$.
    \item Minty at $q^{\star}$: $\langle g(q),\, q - q^{\star}\rangle \leq 0$ for all $q \in X^{\circ}(\mathcal{G})$.
    \item Variational stability at $q^{\star}$: $\langle g(q),\, q - q^{\star}\rangle \leq 0$ for $q$ in a neighborhood of $q^{\star}$, with strict inequality whenever $q \neq q^{\star}$.
\end{itemize}
By Proposition~\ref{lem:reference-invariance}, every inner product of this form is the same for any choice of reference profile. Hence whether $q^{\star}$ (or $g$) satisfies any of these three conditions is determined by the game $\mathcal{G}$ and the strategy profile alone, not by the choice of $(\bar{\sigma}, \bar{\alpha})$ in Definition~\ref{def:PayoffDifferenceField}.

\section{Missing Material of Section \ref{sec:experiments}}\label{sec:Missing_details_experimemts}

\subsection{Treatment Variables and Specification}\label{subsec: TreamentVariables}

At $K=2$ each player's strategy reduces to a $2$-dimensional
vector, so the joint state is
\[
x = (p_L,\, p_H,\, r_L,\, r_H) \in [0,1]^4,
\]
where $p_t$ is the probability that the type-$t$ sender plays the
high signal and $r_m$ is the probability that the receiver plays
the high action on signal $m$. \emph{Local} runs start near the
family's target equilibrium $\sigma^{*}$, with a per-coordinate
Euclidean spread of $\rho \approx 0.07$; \emph{global} runs start
uniformly on the interior of the simplex. All four algorithms share
the same $(\sigma_0,\rho_0)$ at each of the $n=50$ runs
within an init. scheme, so cross-algorithm comparisons are paired. In multi-equilibrium families, $\sigma^*$ refers to the family's fixed reference
  equilibrium (e.g. Riley outcome for the Spence case, pooling at the low signal for the multiple pooling case and the riley outcome for the multiple separating case).

Across all experiments, each seed independently draws the initial
  policies $(\sigma_0, \rho_0)$ per row from
  $\mathrm{Dirichlet}(50\,\sigma^{*}+\mathbf{1})$ for \textsc{local}
  init (concentrated near the target) or $\mathrm{Dirichlet}(\mathbf{1})$
  for \textsc{global} init (uniform on the simplex), and seeds any
  algorithm-internal randomness. The reference profile
  $\sigma^{}_{\mathrm{ref}}$ is the family's designated
  centroid — the unique equilibrium when one exists, and the fixed reference equilibrium above in multi-equilibrium families.
\begin{table}[htp]
\centering
\caption{
Treatment variables in the preferred specification, using the hyperparameter values that delivered the strongest convergence performance in the robustness analysis}
\label{tab:treatments}
\begin{tabular}{p{3.3cm}p{8.6cm}}
\toprule
\textbf{Variable} & \textbf{Value} \\
\midrule
Families & unique separating; unique pooling ($m_0$, $m_1$); pooling and separating; unique semi-separating; multiple pooling; multiple separating (Riley/anti) \\
Algorithms & PGA, OMA, Opt.\,OMA, PPO \\
Step size $\eta$ & PGA $0.3$; OMA / Opt.\,OMA $1.0$; PPO $5\!\times\!10^{-2}$ \\
Step-size Schedule & constant \\
Initialisation regime & local (near target, $\rho\!\approx\!0.07$); global (uniform on simplex) \\
Horizon $T$ & $2000$ (classical); $240$ rounds (PPO) \\
Runs per treatment & $n = 50$ \\
Conv. Tolerance $\delta$ & $10^{-2}$ \\
\bottomrule
\end{tabular}
\end{table}

\subsection{Game parameterisations}
\label{app:families}

We describe the parameterizations of the different equilibrium classes.

\paragraph{Unique separating (honest-signaling)}
  \label{app:family-uniq-sep}

  We define $u_S(\theta, s, a) = \alpha_S \cdot \mathbf{1}[s = \theta] + \varepsilon_S$
  and $u_R(\theta, s, a) = \alpha_R \cdot \mathbf{1}[a = \theta] + \varepsilon_R$,
  where $\alpha_S, \alpha_R$ are sender / receiver reward multiplier for matching behavior. Further,
  $\varepsilon = (\varepsilon_S, \varepsilon_R)$ are bounded perturbations.
  Note that for sufficiently high $\alpha_S, \alpha_R$ relative to the
  perturbation budget, we obtain \emph{sender dominance} (each type
  strictly prefers its own signal, regardless of the receiver) and a
  \emph{receiver unique best response} (the matched action strictly
  beats any alternative on each signal).

  \paragraph{Unique pooling}
  \label{app:family-uniq-pool}

  Given the pooling signal $s_*$, we have $u_S(\theta, s_*, a) = \alpha_S$;
  for all other signals $s \neq s_*$, $u_S(\theta, s, a) = \beta_S < \alpha_S$
  (a strictly worse outside option, independent of $a$). Receiver
  payoff is the matching premium
  $u_R(\theta, s, a) = \alpha_R \cdot \mathbf{1}[a = \theta]$
  (reward iff the action matches the realised type). Under the on-path
  posterior on the pooled signal --- which equals the prior $\pi$ ---
  the receiver's expected utility is the probability that the action and the type match.                    Hence, the best response is unique iff the prior is strictly non-uniform.

  \paragraph{Pooling and separating (Spence-style Job-Market Signaling)}
  \label{app:family-spence}

  We consider a discrete job-market signaling model of
  \citet{spence1978job}: the receiver pays a wage $w_a$ to each
  action, and the sender of type $\theta$ pays signaling cost $c_\theta$
  per unit of high signal. Concretely, for $K=2$,
  $u_S(\theta, s, a) = w_a - c_\theta \cdot \mathrm{1}_{\{s=1\}}$, and the receiver
  earns $+\alpha_R$ when its wage matches the realised type and
  $-\alpha_R$ otherwise, i.e.
  $u_R(\theta, s, a) = \alpha_R \cdot (2\,\mathbf{1}[a = \theta] - 1)$.
  Single-crossing requires the marginal cost to fall in $\theta$:
  $c_L > c_H$.\\
  The single-crossing condition $c_L > c_H$ ensures that a high-signal
  threshold separates the two types in equilibrium; a non-trivial wage
  gap $w_H > w_L$ ensures the receiver actually discriminates. With
  $c_H \le w_H - w_L \le c_L$ and $c_H > 0$, both a
  separating $\PBE$ and a pooling $\PBE$ survive.

  \paragraph{Unique semi-separating (Kreps--Wilson reputation)}
  \label{app:family-reputation}

  We consider a discrete reputation game inspired by
\citet{kreps1982reputation}. The strategic-type sender chooses
  between Accommodate ($s=0$) and Prey ($s=1$); the commitment-type
  sender always plays Prey (we use ``Prey'' throughout, in line with
  the table below); the receiver chooses to STAY ($a=0$) or EXIT
  ($a=1$). Sender payoffs are parameterised by the deterrence-stay
  payoff $D_1$ on Accommodate, the prey-stay payoff $P_1$, the
  maximal payoff $M_1$ from successfully Preying when the receiver
  Exits, and the accommodate-exit payoff $A_1$; receiver payoffs by
  the rate of return $D_2 > 0$ when staying against a Sane sender
  and $P_2 < 0$ when staying against a Crazy one. The parameter
  $\delta > 0$ is the Crazy sender's strict-Prey margin (it makes
  Prey strictly dominant for the Crazy type; it is \emph{not} a
  discount factor and does not enter receiver payoffs), see also the following table.

   \begin{table}[h]                             \label{tab:semiSeparating}
    \centering
    \caption{Stage payoffs of the reputation signalling game per
    sender type; each cell reports $(u_S,\,u_R)$. The \emph{Sane}
    sender (prior $p$) chooses freely between Accommodate and Prey;
    the \emph{Crazy} sender (prior $1-p$) plays Prey by construction
    (it is strictly dominant by $\delta > 0$, with $X$ the Crazy
    baseline payoff). The conditions $M_1 > D_1 > P_1$, $A_1 > P_1$,
    $D_2 > 0 > P_2$ and $pD_2 + (1-p)P_2 > 0$ with $p \in (0,1)$ jointly pin down the unique semi-separating $\PBE$.}
    \label{tab:reputation_payoffs}
    \setlength{\tabcolsep}{6pt}
    \begin{minipage}[t]{0.52\linewidth}
    \textbf{\hspace{91pt} Sane sender (prob.\ $p$)}\\[2pt]
    \begin{tabular}{l|cc}
    \toprule
    & \textsc{Stay} ($a{=}0$) & \textsc{Exit} ($a{=}1$) \\
    \midrule
    Accommodate ($s{=}0$) & $(D_1,\,D_2)$ & $(A_1,\,0)$ \\
    Prey ($s{=}1$)        & $(P_1,\,D_2)$ & $(M_1,\,0)$ \\
    \bottomrule
    \end{tabular}
    \end{minipage}\hfill
    \begin{minipage}[t]{0.44\linewidth}
    \centering
    \textbf{Crazy sender (prob.\ $1-p$)}\\[2pt]
    \begin{tabular}{l|cc}
    \toprule
    & \textsc{Stay} ($a{=}0$) & \textsc{Exit} ($a{=}1$) \\
    \midrule
     & $(X-\delta,\,P_2)$ & $(X-\delta,\,0)$ \\
     & $(X,\,P_2)$        & $(X,\,0)$ \\
    \bottomrule
    \end{tabular}
    \end{minipage}
    \end{table}

  In the unique semi-separating $\PBE$ each player's mixing
probability is fixed by the \emph{other} player's indifference
condition. The Sane sender's mix probability $q^{*}$ on Prey is
pinned by the receiver's indifference between Stay and Exit on the
on-path Prey signal: the posterior on Sane must equal $1-\tau$, with
$\tau \equiv D_2 / (D_2 - P_2)$, which yields
$q^{*} = (1-p)(1-\tau)/(\tau p)$; this lies in $(0,1)$ iff $p > 1-\tau$.
The receiver's mix probability $r^{*}$ on Stay (conditional on the
Prey signal) is pinned by the Sane sender's indifference between
Accommodate, after which the receiver Stays, and Prey, yielding
$r^{*} = (M_1 - D_1)/(M_1 - P_1)$. Neither expression involves $A_1$;
the condition $A_1 > P_1$ only rules out the pure pooling PE.

  \paragraph{Multiple pooling}
  \label{app:family-multi-pool}

  Sender utility depends only on whether the receiver plays the
  prior-best-response action: $u_S(\theta, s, a) = \beta \cdot
  \mathbf{1}[a = a^{*}]$ with $a^{*} = \arg\max_a \sum_\theta \pi(\theta)
  u_R(\theta, s, a)$. Receiver matches signal-conditional type as in
  the pooling family.
  A sufficiently high $\beta$ ensures the sender strictly prefers the
  receiver to play $a^{*}$, which makes both pure pooling profiles
  (pool at $s_0$, pool at $s_1$) potential candidates for equilibria.
  A sufficiently high value for $\alpha_R$ ensures the receiver's
  prior-BR is unique on every signal.

  \paragraph{Multiple separating (Riley + anti-Riley)}
  \label{app:family-multi-sep}

  This class admits two pure-strategy separating $\PBE$ at $K=2$ ---
  Riley ($\sigma^{*}=I$) and anti-Riley ($\sigma^{*}=$ swap).

  \textbf{K=2 instance (used in
  \Cref{tab:endogenous_constant_results}).}
  For each type, sender payoffs depend on whether the realised
  $(s, a)$ pair lies on the Riley or anti-Riley path:
  \begin{align*}
  u_S(\theta_0, s_0, a_0) &= R   & u_S(\theta_0, s_1, a_0) &= A   \\
  u_S(\theta_0, s_1, a_1) &= D_R & u_S(\theta_0, s_0, a_1) &= D_A,
  \end{align*}
  with type $\theta_1$ symmetric under the type/signal/action swap.
  The receiver matches: $u_R(\theta, s, a) = \alpha_R \cdot
  \mathbf{1}[a = \theta]$. The conditions
  $R > A > D_A > D_R \ge 0$ together with a strictly non-uniform prior
  imply that both pure separating profiles are \emph{strict} $\PBE$:
  the Riley IC slack is $R - D_R$ and the anti-Riley IC slack is
  $A - D_A$. (Note that mixed-semi-separating equilibria may still survive in cases where the riley and anti-riley separating equilibrium exist.)
  To rule out pooling, we additionally require the condition:
  \[
   R+D_R > A+D_A.
  \]

Finally, for completeness, we provide the used game parameterizations for all equilibrium classes, which we used in the simulations to generate the results that are illustrated in Table \ref{tab:endogenous_constant_results}.

  \input{family_params_table}

For completeness, we provide the game parameterizations for all games in the $7$ different classes. The game parameters used in Table \ref{tab:endogenous_constant_results} are stated in Table \ref{tab:family-params}.

\subsection{Algorithm details and hyperparameters}
\label{app:algorithms}

We specify the four algorithms and the hyperparameters used in
  every cell of \Cref{tab:endogenous_constant_results}. All step
  sizes are constant ($\eta_S = \eta_R = \eta_0$, no decay schedule).

All four algorithms operate on the same simplex-valued strategies. The three classical learners (PGA, OMA,    Opt.\,OMA) consume full-information feedback from a shared oracle.          PPO is the only algorithm that does \emph{not} use this oracle:                        it samples episodes from the joint distribution induced by                             $(\sigma, \rho, \text{prior})$, where $\sigma, \rho$ describe the sender and the receiver (mixed) strategy respectively.

\paragraph{Feedback oracle.}
  With $\sigma[\tau,m]=q_S(m\mid\tau)$ and $\rho[m,a]=q_R(a\mid m)$, the oracle returns
  \begin{equation}\label{eq:oracle}
    Q_S[\tau,m]=\sum_{a\in A}\rho[m,a]\,u_S(\tau,m,a),\qquad
    Q_R[m,a]=\sum_{\tau\in\Theta}\mu(\tau\mid m)\,u_R(\tau,m,a),
  \end{equation}
  where $\mu$ is the posterior \eqref{eq. posterior update}. On $X^{\circ}(\mathcal{G})$,
  $Q_S$ and $Q_R$ coincide with $g_S$ and $g_R$ in
  \eqref{eq:g-sender}--\eqref{eq:g-receiver} up to an additive constant in each row,
  to which all three updates below are invariant. If a signal $m$ has zero
  probability under $(\pi,\sigma)$, the oracle sets $\mu(\cdot\mid m):=\pi$.

  \subsubsection{Projected Gradient Ascent (PGA)}
  \label{app:algo-pga}
  Per-row Euclidean ascent followed by exact simplex projection
  \citep{wang2013simplex}:
  \[
  \sigma_{t+1}[\tau,:] = \Pi_{\Delta_K}\!\bigl(\sigma_t[\tau,:] +
  \eta_0\,Q_S[\tau,:]\bigr),\qquad
  \rho_{t+1}[m,:] = \Pi_{\Delta_K}\!\bigl(\rho_t[m,:] + \eta_0\,Q_R[m,:]\bigr).
  \]                                      Note that $Q_S, Q_R$ are the full-information feedback gradients from \ref{eq:oracle}. We use the step size: $\eta_0 = 0.3$.

  \subsubsection{Entropic Online Mirror Ascent (OMA)}
  \label{app:algo-oma}
  Next, we have
  \[
  \sigma_{t+1}[\tau,:] \propto
  \sigma_t[\tau,:]\,\exp\!\bigl(\eta_0\,Q_S[\tau,:]\bigr),\qquad
  \rho_{t+1}[m,:] \propto \rho_t[m,:]\,\exp\!\bigl(\eta_0\,Q_R[m,:]\bigr),
  \]
  with row-wise normalisation. We subtract the row-wise maximum of
  $\eta_0\,Q$ before exponentiation for numerical stability. Iterates
  remain strictly in the relative interior of the simplex. Step
  size: $\eta_0 = 1.0$.

  \subsubsection{Optimistic Entropic OMA (\texorpdfstring{Opt.\,OMA}{Opt. OMA})}
  \label{app:algo-opt-oma}
  Two iterates per player as in \citet{syrgkanis2015fast}: a pivot
  $\bar\sigma_t$ and a played strategy $\sigma_t$. Per round, observe
  gradients at the played strategies, take an entropic step on the
  pivot, and an extra prediction step on top of the new pivot:
  \[
  \bar\sigma_{t+1}[\tau,:] \propto \bar\sigma_t[\tau,:]\,\exp(\eta_0\,Q_S[\tau,:]),\qquad
  \sigma_{t+1}[\tau,:] \propto \bar\sigma_{t+1}[\tau,:]\,\exp(\eta_0\,Q_S[\tau,:]),
  \]
  and analogously for the receiver. Step size: $\eta_0 = 1.0$.

  \subsubsection{Proximal Policy Optimisation (PPO)}
  \label{app:algo-ppo}

  \paragraph{Episode structure.}
  PPO sees a \emph{single-step episodic environment}: each episode
  is one play of the signaling game, $(t, m, a) \sim
  \mathrm{prior}\cdot\sigma\cdot\rho$, and the per-agent reward is
  the realised one-shot payoff. Because each episode is one transition,
  neither a discount factor $\gamma$ nor a GAE bootstrap parameter
  $\lambda_{\text{GAE}}$ applies: the per-episode return is the
  realised reward and the advantage is the one-step residual, for more details, see             \citep{schulman2017ppo}.

  \paragraph{Tabular policies and values.}
  Each agent has tabular logits $\theta\in\mathbb{R}^{K\times K}$
  with $\pi(\cdot \mid s) = \mathrm{softmax}(\theta[s,:])$ and a
  tabular value head $V \in \mathbb{R}^{K}$ over the same state set,
  which matches the function-approximation class to OMA's policy
  class.

  \paragraph{Round.}
  Each PPO round (i) collects one batch of $N_{\text{rollout}}=512$ joint episodes per   agent under the current policies, shared by both agents (ii) computes one-step advantages
  $A = r - V(s)$, normalised to zero mean and unit variance over the
  batch; (iii) runs $E=4$ epochs of mini-batch
  ($M_{\text{batch}}=128$) clipped-surrogate updates per agent with
  \[
  \mathcal{L}
  = -\,\mathbb{E}\!\Bigl[\min\!\bigl(\, r_\theta\, \widehat A,\ \mathrm{clip}(r_\theta,
  1{-}\epsilon, 1{+}\epsilon)\,\widehat A\bigr)\Bigr]
  + c_v\,\mathrm{MSE}\bigl(V(s), r\bigr)
  - c_e\,H(\pi_\theta),
  \]
  where $r_\theta = \pi_\theta(a\mid s)/\pi_{\theta_{\text{old}}}(a\mid s)$
  is the importance-sampling ratio and $H(\pi_\theta)$ the per-state
  entropy. The two agents are updated independently on this shared batch.

  \paragraph{Hyperparameters.}
  \Cref{tab:ppo-hyperparams} lists every PPO hyperparameter used in
  the main cross-tab.

  \begin{table}[h]
  \centering
  \small
  \caption{PPO hyperparameters used in every PPO cell of
  \Cref{tab:endogenous_constant_results}. Discount factor $\gamma$
  and GAE parameter $\lambda_{\text{GAE}}$ are not applicable because
  each episode is a single transition; the advantage is the one-step
  residual $r - V(s)$.}
  \label{tab:ppo-hyperparams}
  \begin{tabular}{l l l}
  \toprule
  Symbol & Value & Description \\
  \midrule
  $N_{\text{rollout}}$ & $512$  & Episodes per round (per agent) \\
  $E$                  & $4$    & Optimisation epochs per round \\
  $M_{\text{batch}}$   & $128$  & Mini-batch size \\
  $\epsilon$           & $0.2$  & Clip ratio (clipped surrogate) \\
  $c_v$                & $0.5$  & Value-loss coefficient \\
  $c_e$                & $0.01$ & Entropy-bonus coefficient \\
  $\eta_0$             & $5\!\times\!10^{-2}$ & Adam learning rate \\
  $g_{\max}$           & $0.5$  & Global gradient-norm clip \\
  Optimiser            & Adam   & PyTorch defaults: $\beta_1=0.9, \beta_2=0.999,
  \epsilon_{\text{Adam}}=10^{-8}$ \\
  Advantage normalisation & on  & per-batch zero-mean, unit-variance \\
  $T_{\text{PPO}}$     & $240$ rounds & main cross-tab horizon \\
  \bottomrule
  \end{tabular}
  \end{table}

\section{Standard deviations}
\label{app:stats}

\Cref{tab:endogenous_constant_results} reports
point estimates over $n=50$ paired-init seeds per cell.
\Cref{tab:endogenous_constant_stds} reports the matching
across-seed sample standard deviations for each metric.
$\mathrm{sd}(\textsc{Conv.})$ is the sample standard deviation of
the binary convergence indicator (so $\mathrm{sd}(\textsc{Conv.})
\approx \sqrt{p(1-p)}$ for $\textsc{Conv.}=p$);
$\mathrm{sd}(\textsc{Dist.})$ is the sample standard deviation of
the terminal distance over all 50 runs; and $\mathrm{sd}(\textsc{Hit})$
is the sample standard deviation of the hitting time over the
converged subset (``$-$'' when no run converged or only one did).

\begin{table}[t]
\centering
\caption{Across-seed sample standard deviations companion to the
main results table. Each cell aggregates $n=50$ paired-init seeds.
$\textsc{Hit}$ standard deviations are computed over the converged
subset only.}
\label{tab:endogenous_constant_stds}
\setlength{\tabcolsep}{5pt}
\footnotesize
\begin{tabular}{
  @{}
  ll
  S[table-format=1.3]
  S[table-format=1.3]
  S[table-format=2.1]
  S[table-format=1.3]
  S[table-format=1.3]
  S[table-format=2.1]
  @{}
}
\toprule
\multicolumn{2}{c}{} &
\multicolumn{3}{c}{\textbf{Local initialization}} &
\multicolumn{3}{c}{\textbf{Global initialization}} \\
\cmidrule(lr){3-5}
\cmidrule(lr){6-8}
\textbf{Eq. class} &
\textbf{Alg.} &
{\textbf{sd(\textsc{Conv.})}} &
{\textbf{sd(\textsc{Dist.})}} &
{\textbf{sd(\textsc{Hit})}} &
{\textbf{sd(\textsc{Conv.})}} &
{\textbf{sd(\textsc{Dist.})}} &
{\textbf{sd(\textsc{Hit})}} \\
\midrule

Unique separating
  & PGA       & 0.000 & 0.000 & 0.0 & 0.000 & 0.000 & 1.1 \\
  & OMA       & 0.000 & 0.000 & 0.4 & 0.000 & 0.000 & 1.1 \\
  & Opt.\,OMA & 0.000 & 0.000 & 0.0 & 0.000 & 0.000 & 1.0 \\
  & PPO       & 0.000 & 0.000 & 0.8 & 0.000 & 0.000 & 1.7 \\
\midrule

Unique pooling ($m_0$)
  & PGA       & 0.000 & 0.000 & 0.2 & 0.000 & 0.000 & 5.5 \\
  & OMA       & 0.000 & 0.000 & 1.9 & 0.000 & 0.000 & 6.0 \\
  & Opt.\,OMA & 0.000 & 0.000 & 1.6 & 0.000 & 0.000 & 4.8 \\
  & PPO       & 0.000 & 0.000 & 3.4 & 0.000 & 0.000 & 8.0 \\
\midrule

Unique pooling ($m_1$)
  & PGA       & 0.000 & 0.000 & 0.2 & 0.000 & 0.000 & 5.4 \\
  & OMA       & 0.000 & 0.000 & 1.7 & 0.000 & 0.000 & 6.3 \\
  & Opt.\,OMA & 0.000 & 0.000 & 1.4 & 0.000 & 0.000 & 5.5 \\
  & PPO       & 0.141 & 0.002 & 3.1 & 0.000 & 0.000 & 8.3 \\
\midrule

Pooling and separating
  & PGA       & 0.000 & 0.000 & 0.0 & 0.000 & 0.000 & 4.4 \\
  & OMA       & 0.000 & 0.000 & 0.6 & 0.000 & 0.000 & 3.3 \\
  & Opt.\,OMA & 0.000 & 0.000 & 0.3 & 0.000 & 0.000 & 2.9 \\
  & PPO       & 0.000 & 0.000 & 1.0 & 0.000 & 0.000 & 7.7 \\
\midrule

Unique semi-separating
  & PGA       & 0.000 & 0.079 & {\text{--}} & 0.000 & 0.094 & {\text{--}} \\
  & OMA       & 0.000 & 0.198 & {\text{--}} & 0.000 & 0.234 & {\text{--}} \\
  & Opt.\,OMA & 0.000 & 0.000 & 10.0       & 0.000 & 0.000 & 41.0 \\
  & PPO       & 0.000 & 0.203 & {\text{--}} & 0.000 & 0.220 & {\text{--}} \\
\midrule

Multiple pooling
  & PGA       & 0.501 & 0.026 & 1.5  & 0.388 & 0.205 & 7.8 \\
  & OMA       & 0.505 & 0.027 & 34.5 & 0.388 & 0.147 & 10.3 \\
  & Opt.\,OMA & 0.505 & 0.027 & 8.3  & 0.370 & 0.152 & 22.1 \\
  & PPO       & 0.328 & 0.222 & 48.2 & 0.141 & 0.005 & 27.3 \\
\midrule

Multi-separating (Riley/anti)
  & PGA       & 0.000 & 0.000 & 0.0 & 0.000 & 0.000 & 3.4 \\
  & OMA       & 0.000 & 0.000 & 0.7 & 0.000 & 0.000 & 5.6 \\
  & Opt.\,OMA & 0.000 & 0.000 & 0.4 & 0.000 & 0.000 & 4.8 \\
  & PPO       & 0.000 & 0.001 & 0.9 & 0.000 & 0.000 & 2.1 \\

\bottomrule
\end{tabular}
\end{table}

Table \ref{tab:endogenous_constant_stds} makes the qualitative interpretation precise. Cells with
  $\textsc{Conv.} \in \{0.000,\,1.000\}$ in the main Table \ref{tab:endogenous_constant_results} also
  have $\mathrm{sd}(\textsc{Conv.}) = 0$ (every seed had the same
  outcome). Across-seed variability in $\textsc{Conv.}$ is non-trivial
  in two locations:
  (i) the entire multi-pooling family
  ($\mathrm{sd}(\textsc{Conv.}) \in [0.141,\,0.505]$ across the eight
  algorithm$\times$init cells), which is the only family where
  convergence is genuinely seed-dependent for every algorithm; and ii) PPO under local initialisation on unique pooling at $m_1$
  ($\mathrm{sd}(\textsc{Conv.}) = 0.141$, consistent with $1$ out of
  $50$ seeds failing).
  The remaining $47$ of the $56$ algorithm$\times$init$\times$family
  cells in \Cref{tab:endogenous_constant_stds} have
  $\mathrm{sd}(\textsc{Conv.}) = 0$, so the qualitative reading
  ``either every seed converges or every seed fails'' holds outside
  those two locations. The $\mathrm{sd}(\textsc{Hit})$ entries are
  heavy-tailed in multi-pooling (e.g.\ OMA local at $34.5$ rounds
  and PPO local at $48.2$, against medians of $11.5$ and $29.5$
  respectively).

  \subsection{Experiments compute resources}\label{subsec:ComputeResources}

The experiments/simulations are small-scale. In other words, no additional computational resources are required and the experiments are feasible on a local machine in a couple of hours (e.g. in our case, all experiments were run on a 14" Apple MacBook Pro (Nov. 2024) with an Apple M4 Pro processor and 48 GB of unified memory, running macOS Tahoe 26.3.1 (a).)

\section{Missing Material of Section \ref{sec:Beyond K=2}}

\subsection{Four-type benchmark games}
\label{app:k4-benchmarks}

We extend the unique separating and unique semi-separating benchmarks to
$K=4$. Throughout this section,
$\Theta=\{\theta_1,\ldots,\theta_4\}$,
$\Sigma=\{s_1,\ldots,s_4\}$, and $A=\{a_1,\ldots,a_4\}$.
Here, we write $Q_S=(q_S(s_j\mid\theta_i))_{i,j=1}^4$ and
$Q_R=(q_R(a_j\mid s_i))_{i,j=1}^4$ for the Sender and Receiver strategy
matrices, respectively. The constructions below admit a unique PE among all
behavioral strategy profiles. In both cases every signal is on-path at the
equilibrium, so the equilibrium belief system is uniquely determined by
Bayes' rule.

\subsubsection{Unique separating (honest-signaling game)}
\label{app:k4-honest}

We retain the matching rewards of the binary honest-signaling benchmark.
For $i,j,\ell\in\{1,\ldots,4\}$, define
\begin{align}
u_S(\theta_i,s_j,a_\ell)
&=\alpha_S\mathbf{1}\{j=i\}
  +\varepsilon_S(\theta_i,s_j,a_\ell),
\label{eq:k4-honest-sender}\\
u_R(\theta_i,s_j,a_\ell)
&=\alpha_R\mathbf{1}\{\ell=i\}
  +\varepsilon_R(\theta_i,s_j,a_\ell).
\label{eq:k4-honest-receiver}
\end{align}
The perturbations are fixed payoff functions satisfying
$|\varepsilon_S(\theta_i,s_j,a_\ell)|\le b_S$ and
$|\varepsilon_R(\theta_i,s_j,a_\ell)|\le b_R$, where $b_S,b_R\ge0$.
We assume a full-support prior and
\begin{equation}
\alpha_S>2b_S,
\qquad
\alpha_R>2b_R.
\label{eq:k4-honest-conditions}
\end{equation}
Under these bounds, truthful signaling is strictly dominant. The Receiver
uniquely matches the revealed type.

\begin{proposition}[Unique separating PE]
\label{prop:k4-honest-unique}
Under \eqref{eq:k4-honest-conditions}, the honest-signaling game has a unique
PE. It is the strict Riley PE, given by
\begin{equation}
q_S^\star(s_j\mid\theta_i)=\mathbf{1}\{j=i\},
\qquad
q_R^\star(a_\ell\mid s_j)=\mathbf{1}\{\ell=j\},
\qquad
\mu^\star(\theta_i\mid s_j)=\mathbf{1}\{i=j\}.
\label{eq:k4-honest-equilibrium}
\end{equation}
\end{proposition}

\begin{proof}
Fix a type $\theta_i$ and a signal $s_j\ne s_i$. By
\eqref{eq:k4-honest-conditions},
\[
\min_{a\in A}u_S(\theta_i,s_i,a)
\ge\alpha_S-b_S
> b_S
\ge\max_{a\in A}u_S(\theta_i,s_j,a).
\]
Hence, $s_i$ is strictly dominant for type $\theta_i$, and every PE must have
$q_S(s_i\mid\theta_i)=1$. The prior has full support, so every signal is
on-path and Bayes' rule gives
$\mu(\theta_i\mid s_j)=\mathbf{1}\{i=j\}$. At signal $s_i$, for every
$a_\ell\ne a_i$,
\[
u_R(\theta_i,s_i,a_i)-u_R(\theta_i,s_i,a_\ell)
\ge\alpha_R-2b_R>0.
\]
Therefore the Receiver must choose $a_i$ with probability one. These strict
inequalities also establish that the  assessment is a strict PE,
completing both existence and uniqueness.
\end{proof}

\paragraph{Concrete parameterization.}
For the experiments, we take the uniform prior $\pi(\theta_i)=1/4$ and the
fixed utilities
\begin{equation}
u_S(\theta_i,s_j,a_\ell)
=2\,\mathbf{1}\{j=i\}
u_R(\theta_i,s_j,a_\ell)=2\,\mathbf{1}\{\ell=i\}.
\label{eq:k4-honest-numeric}
\end{equation}
This sets $\alpha_S=\alpha_R=2$, and $b_S=b_R=0$, so
\eqref{eq:k4-honest-conditions} holds.

\subsubsection{Unique semi-separating (multi-threat reputation game)}
\label{app:k4-reputation}

We extend the binary reputation game of \citet{kreps1982reputation} by
introducing three distinct types. Type $\theta_1$ is Sane and
strategic, whereas $\theta_j$, $j\in\{2,3,4\}$, is committed to the
corresponding Prey signal $s_j$. Signal $s_1$ is Accommodate. The Receiver
chooses \textsc{Stay} (labeled $a_1$) or one of three threat-specific
\textsc{Exit} actions ($a_2,a_3,a_4$).

Write $p=\pi(\theta_1)$ (assuming $p \in (0,1)$) and $\pi_j=\pi(\theta_j)>0$ for $j=2,3,4$, with
$p+\pi_2+\pi_3+\pi_4=1$. The Sane Sender's payoffs are
\begin{equation}
u_S(\theta_1,s,a)=
\begin{cases}
D_1, & s=s_1,\ a=a_1,\\
A_1, & s=s_1,\ a\ne a_1,\\
P_1, & s\ne s_1,\ a=a_1,\\
M_1, & s\ne s_1,\ a\ne a_1.
\end{cases}
\label{eq:k4-reputation-sane}
\end{equation}
For each type $j\in\{2,3,4\}$, set
\begin{equation}
u_S(\theta_j,s,a)=X-\delta\mathbf{1}\{s\ne s_j\},
\qquad \delta>0,
\label{eq:k4-reputation-commitment}
\end{equation}
where $X\in\mathbb R$ is a baseline payoff. Thus, each type
strictly prefers its designated Prey signal regardless of the Receiver's
action. Receiver payoffs are independent of the signal and are specified in
Table~\ref{tab:k4-reputation-receiver}. The appropriate Exit action gives
payoff zero, while an Exit action tailored to a different type
incurs a penalty $\zeta>0$.

\begin{table}[t]
\centering
\caption{Receiver payoffs in the four-type reputation game, for every
$s\in\Sigma$. Action $a_j$ is the Exit response tailored to
type $\theta_j$, $j\in\{2,3,4\}$.}
\label{tab:k4-reputation-receiver}
\begin{tabular}{lcccc}
\toprule
Type & \textsc{Stay} ($a_1$) & \textsc{Exit} ($a_2$)
     & \textsc{Exit} ($a_3$) & \textsc{Exit} ($a_4$)\\
\midrule
$\theta_1$ (Sane) & $D_2$ & $0$ & $0$ & $0$\\
$\theta_2$ & $P_2$ & $0$ & $-\zeta$ & $-\zeta$\\
$\theta_3$ & $P_2$ & $-\zeta$ & $0$ & $-\zeta$\\
$\theta_4$ & $P_2$ & $-\zeta$ & $-\zeta$ & $0$\\
\bottomrule
\end{tabular}
\end{table}

We impose the same payoff ordering and prior condition as in the binary
reputation benchmark, together with a strictly positive penalty for an
inappropriate Exit response:
\begin{equation}
\begin{gathered}
M_1>D_1>P_1,\qquad A_1>P_1,\qquad D_2>0>P_2,\\
\delta>0,\qquad \zeta>0,\qquad
pD_2+(1-p)P_2>0.
\end{gathered}
\label{eq:k4-reputation-conditions}
\end{equation}
Define
\begin{equation}
b_j:=\frac{-\pi_jP_2}{pD_2}\quad(j=2,3,4),
\qquad b:=\sum_{j=2}^4b_j,
\qquad r^\star:=\frac{M_1-D_1}{M_1-P_1}.
\label{eq:k4-reputation-mixing}
\end{equation}
The conditions in \eqref{eq:k4-reputation-conditions} imply
$b_j>0$, $b<1$, and $r^\star\in(0,1)$.

\begin{proposition}[Unique semi-separating PE]
\label{prop:k4-reputation-unique}
Under \eqref{eq:k4-reputation-conditions}, the four-type reputation game
has a unique PE. Its strategy matrices are
\begin{equation}
Q_S^\star=
\begin{pmatrix}
1-b & b_2 & b_3 & b_4\\
0 & 1 & 0 & 0\\
0 & 0 & 1 & 0\\
0 & 0 & 0 & 1
\end{pmatrix},
\qquad
Q_R^\star=
\begin{pmatrix}
1 & 0 & 0 & 0\\
r^\star & 1-r^\star & 0 & 0\\
r^\star & 0 & 1-r^\star & 0\\
r^\star & 0 & 0 & 1-r^\star
\end{pmatrix}.
\label{eq:k4-reputation-equilibrium}
\end{equation}
The equilibrium is semi-separating: the Sane type mixes over all four
signals, and each Prey signal is shared with its associated type.
\end{proposition}

\begin{proof}
We first derive the strategy profile of any PE. By
\eqref{eq:k4-reputation-commitment}, each type $\theta_j$, $j\ge2$, must
play $s_j$ with probability one. Hence every Prey signal is on-path. Write
$x_j=q_S(s_j\mid\theta_1)$. Bayes' rule gives
\[
\mu(\theta_1\mid s_j)=\frac{p x_j}{p x_j+\pi_j},
\qquad
\mu(\theta_j\mid s_j)=\frac{\pi_j}{p x_j+\pi_j}.
\]
All other types have posterior probability zero. At $s_j$, the appropriate
Exit action $a_j$ gives expected payoff zero. Each other Exit action has
strictly negative expected payoff
$-\zeta\pi_j/(p x_j+\pi_j)$, while the expected payoff from Stay is
\begin{equation}
\frac{p x_jD_2+\pi_jP_2}{p x_j+\pi_j}.
\label{eq:k4-reputation-stay-value}
\end{equation}
Consequently, Stay is uniquely optimal if $x_j>b_j$, the appropriate Exit
is uniquely optimal if $x_j<b_j$, and precisely these two actions are best
responses if $x_j=b_j$.

Suppose $x_j>b_j$ for some $j$. Then $x_j>0$ and the Receiver must Stay at
$s_j$, giving the Sane Sender payoff $P_1$. However, deviating to
Accommodate guarantees at least $\min\{D_1,A_1\}>P_1$, independently of
the Receiver's response there. This contradicts Sender optimality.
Therefore $x_j\le b_j$ for every $j$. Since $b<1$, it follows that
\[
q_S(s_1\mid\theta_1)=1-\sum_{j=2}^4x_j\ge1-b>0.
\]
Only the Sane type sends $s_1$, so its posterior probability there is one
and the Receiver uniquely chooses Stay. Hence the Sane Sender's equilibrium
payoff is $D_1$.

If $x_j<b_j$ for any $j$, the appropriate Exit action is uniquely optimal
at $s_j$, and the Sane Sender can obtain $M_1>D_1$ by choosing that signal.
Thus $x_j=b_j$ for every $j$. Each Prey signal is therefore used with
positive probability by the Sane type. Its indifference between
Accommodate and $s_j$ requires
\[
r_jP_1+(1-r_j)M_1=D_1,
\qquad r_j:=q_R(a_1\mid s_j),
\]
which uniquely gives $r_j=r^\star$. No other Exit action can receive
positive probability, so every PE must have the strategy matrices in
\eqref{eq:k4-reputation-equilibrium}.

To verify existence, equip this profile with its Bayes beliefs. At $s_1$
the Receiver uniquely prefers Stay. At each $s_j$, $j\ge2$,
\begin{equation}
\mu^\star(\theta_1\mid s_j)=\frac{-P_2}{D_2-P_2},
\qquad
\mu^\star(\theta_j\mid s_j)=\frac{D_2}{D_2-P_2},
\label{eq:k4-reputation-beliefs}
\end{equation}
and every other type has posterior probability zero. Stay and $a_j$ both
give expected payoff zero, while the other actions give strictly negative
payoffs. The Sane Sender obtains $D_1$ from all four signals, and each
commitment type strictly prefers its own signal. Sequential rationality
therefore holds. Finally, all signals are on-path, so the belief system is
unique as well.
\end{proof}

\paragraph{Concrete parameterization.}
We retain the binary reputation benchmark's numerical payoffs and split
its commitment-type prior equally among the three commitment types:
\begin{equation}
\begin{gathered}
(\pi(\theta_1),\pi(\theta_2),\pi(\theta_3),\pi(\theta_4))
=\left(\frac35,\frac{2}{15},\frac{2}{15},\frac{2}{15}\right),\\
(D_1,P_1,M_1,A_1)=(2,1,4,2),\qquad
(D_2,P_2)=(1,-1),\qquad X=\delta=\zeta=1.
\end{gathered}
\label{eq:k4-reputation-numeric}
\end{equation}
The prior condition holds since
$pD_2+(1-p)P_2=1/5>0$. Here $b_j=2/9$, $b=2/3$, and
$r^\star=2/3$, so the exact target matrices are
\begin{equation}
Q_S^\star=
\begin{pmatrix}
\frac13 & \frac29 & \frac29 & \frac29\\
0 & 1 & 0 & 0\\
0 & 0 & 1 & 0\\
0 & 0 & 0 & 1
\end{pmatrix},
\qquad
Q_R^\star=
\begin{pmatrix}
1 & 0 & 0 & 0\\
\frac23 & \frac13 & 0 & 0\\
\frac23 & 0 & \frac13 & 0\\
\frac23 & 0 & 0 & \frac13
\end{pmatrix}.
\label{eq:k4-reputation-numeric-target}
\end{equation}
After Accommodate, the Sane posterior is one; after each Prey signal,
the posterior assigns probability $1/2$ to the Sane type and $1/2$ to the
associated commitment type. The signal probabilities are
$(1/5,4/15,4/15,4/15)$, and every Receiver action occurs with positive
probability. The above target profile is used for local initialization
and distance-to-equilibrium evaluation.

\end{document}

%% file: cross_tab.tex
\begin{table}[t]
\centering
\caption{Aggregate results ($T=2000$) for one benchmark game per family (Table~\ref{tab:family-params}), with preferred constant step sizes.}
\label{tab:endogenous_constant_results}
\setlength{\tabcolsep}{4pt}
\begin{tabular}{l cccc cccc}
\toprule
 & \multicolumn{4}{c}{Local init.} & \multicolumn{4}{c}{Global init.} \\
\cmidrule(lr){2-5}\cmidrule(lr){6-9}
 & PGA & OMA & Opt.\,OMA & PPO & PGA & OMA & Opt.\,OMA & PPO \\
\midrule
\multicolumn{9}{l}{\textsc{Hit} --- full convergence classes$^{\dagger}$} \\
Unique separating          & 1 & 1 & 1 & 3              & 4   & 5  & 4  & 11 \\
Unique pooling ($m_0$)     & 1 & 2 & 1 & 5              & 9.5 & 13 & 12 & 20 \\
Unique pooling ($m_1$)     & 1 & 2 & 1 & 5$^{\dagger}$  & 8   & 13 & 11 & 21 \\
Pooling \& separating      & 1 & 2 & 1 & 4              & 4   & 6  & 5  & 15 \\
Mult.\ sep.                & 1 & 2 & 1 & 3              & 6   & 7  & 6  & 10 \\
\midrule
\multicolumn{9}{l}{Partial convergence classes} \\
Semi-sep.\ \textsc{Conv.}    & .000 & .000 & 1.000 & .000 & .000 & .000 & 1.000 & .000 \\
Semi-sep.\ \textsc{Dist.}    & .460 & .853 & .000  & .979 & .440 & .820 & .000  & .890 \\
Semi-sep.\ \textsc{Hit}      & --   & --   & 40.5  & --   & --   & --   & 70    & --   \\
Mult.\ pool.\ \textsc{Conv.} & .560 & .520 & .480  & .880 & .820 & .820 & .840  & .980 \\
Mult.\ pool.\ \textsc{Dist.} & .019 & .020 & .020  & .060 & .082 & .055 & .053  & .001 \\
Mult.\ pool.\ \textsc{Hit}   & 3    & 11.5 & 9     & 29.5 & 10   & 16   & 17.5  & 16   \\
\bottomrule
\multicolumn{9}{l}{\footnotesize $^{\dagger}$ Exception: PPO, Unique pooling ($m_1$), local init.: \textsc{Conv.}$=0.98$.}
\end{tabular}
\end{table}

%% file: family_params_table.tex
\begingroup
  \setlength{\tabcolsep}{4pt}
  \footnotesize
  \begin{longtable}{@{}l l p{0.55\linewidth}@{}}
  \caption{Exact numeric values used to instantiate every benchmark
  game in \Cref{tab:endogenous_constant_results}.}
  \label{tab:family-params} \\
  \toprule
  \textbf{Family} & \textbf{Parameter} & \textbf{Value} \\
  \midrule
  \endfirsthead
  \multicolumn{3}{l}{\emph{(continued from previous page)}} \\
  \toprule
  \textbf{Family} & \textbf{Parameter} & \textbf{Value} \\
  \midrule
  \endhead
  \midrule
  \multicolumn{3}{r}{\emph{(continued on next page)}} \\
  \endfoot
  \bottomrule
  \endlastfoot

  \multirow{2}{*}{\shortstack[l]{\texttt{unique\_separating}\\\texttt{\_honest\_signaling
  }}}
   & prior $(\mu_0, \mu_1)$ & $(0.5,\ 0.5)$ \\
   & $\alpha_S,\ \alpha_R$ & $2.0,\ 2.0$ \\
  \midrule

  \multirow{4}{*}{\shortstack[l]{\texttt{unique\_pooling}}}
   & pool signal $s_*$ & $0$ (\textsc{low}) or $1$ (\textsc{high}) \\
   & prior $(\mu_0, \mu_1)$ & $(0.6,\ 0.4)$ \\
   & $\alpha_S,\ \beta_S$ & $2.0,\ 0.5$ \\
   & $\alpha_R$ & $2.0$ \\
  \midrule

  \multirow{4}{*}{\shortstack[l]{\texttt{spence\_signaling}}}
   & prior $(\mu_L, \mu_H)$ & $(0.4,\ 0.6)$ \\
   & wages $(w_L, w_H)$ & $(0.0,\ 2.0)$ \\
   & costs $(c_L, c_H)$ & $(3.0,\ 1.0)$ \\
   & $\alpha_R$ & $2.0$ \\
  \midrule

  \multirow{4}{*}{\shortstack[l]{\texttt{reputation}\\\texttt{\_signaling}}}
   & sane prior $p$ & $0.6$ \\
   & $D_1, P_1, M_1, A_1$ & $2.0,\ 1.0,\ 4.0,\ 2.0$ \\
   & $D_2, P_2$ & $1.0,\ -1.0$ \\
   & $\delta$ (Crazy dominance margin) & $1.0$ \\
  \midrule

  \multirow{2}{*}{\texttt{multiple\_pooling}}
   & prior $(\mu_0, \mu_1)$ & $(0.6,\ 0.4)$ \\
   & $\beta,\ \alpha_R$ & $1.0,\ 1.0$ \\
  \midrule

  \multirow{4}{*}{\shortstack[l]{\texttt{multiple}\\\texttt{\_separating}}}
   & prior $(\mu_0, \mu_1)$ & $(0.6,\ 0.4)$ \\
   & $R,\ A$ (Riley, anti payoffs) & $4.0,\ 2.0$ \\
   & $D_R,\ D_A$ (deviation payoffs) & $0.0,\ 1.0$ \\
   & $\alpha_R$ & $1.0$ \\

  \end{longtable}
  \endgroup